\documentclass[12pt]{iopart}
\pdfoutput=1
\usepackage{iopams} 
\usepackage{graphicx}
\usepackage{epsfig}
\usepackage{float}
\usepackage{amssymb}
\usepackage{ctable}
\usepackage{url}

\newcommand{\vx}{{\bf x}}
\newcommand{\fnl}{f_{\rm NL}}
\newcommand{\vk}{{\bf k}}

\def \MSUN{M_{\odot}}

\begin{document}
\title[N-body simulations with generic non-Gaussian initial conditions I]{N-body simulations with generic non-Gaussian initial conditions I: Power Spectrum and halo mass function}
\author{Christian Wagner$^{1}$, Licia Verde$^{1,2}$, Lotfi Boubekeur$^{3,4}$\\
$^1$ {\it ICCUB-IEEC, University of Barcelona, Barcelona 08028, Spain.}\\
$^2$ {\it ICREA, Instituci\'o Catalana de Recerca i Estudis Avan\c{c}ats.}\\
$^3$ {\it Departament de F\'isica Te\`orica, Universitat de Val\`encia, E-46100, Burjassot, Spain.}\\
$^4$ {\it Instituto de F\'isica Corpuscular (IFIC),
Universitat de Val\`encia-CSIC, Edificio de Institutos de Paterna,
Apt. 22085, E-46071, Valencia, Spain.}
}
\begin{abstract}
We  address the issue of setting up generic non-Gaussian initial conditions for N-body simulations.  We consider inflationary-motivated primordial non-Gaussianity  where the perturbations in the  Bardeen potential are   given by a dominant Gaussian part plus a non-Gaussian part  specified by its bispectrum.  The approach we explore here  is suitable for any bispectrum, i.e.~it does not have to be of the so-called  separable or factorizable form.   The procedure of generating a non-Gaussian field with a given bispectrum (and a given power spectrum for the Gaussian component) is not univocal, and care must be 
taken so that higher-order corrections do not leave a too large signature on the power 
spectrum. This is so far a limiting factor of our approach.
We then run N-body simulations for  the most popular inflationary-motivated non-Gaussian shapes. The halo mass function and the non-linear power spectrum agree with  theoretical analytical 
approximations proposed in the literature, even if they were so far developed and tested only for a particular shape (the local one). We plan to make the simulations outputs available to the community via the non-Gaussian simulations comparison project web site \url{http://icc.ub.edu/~liciaverde/NGSCP.html}. 
\end{abstract}

\section{Introduction}

The leading theory for the origin of primordial perturbations is inflation. Along with the shape  and amplitude of the primordial power spectrum and the signature of a stochastic background of gravity waves, primordial non-Gaussianity offers a window to probe inflation. 
Even the simplest inflationary models predict  deviations from Gaussian initial conditions, arising from the (self)-interaction of the field during inflation; for the simplest, slow roll, single field model  these deviations  are expected to be unmeasurably small \cite{ABMR,maldacena:2003}.  For a thorough review of inflationary  non-Gaussianity see e.g., \cite{BKMR04} and references therein.

An important  development of the past few years is the realization that  a large, potentially detectable amount of non-Gaussianity  can be produced by inflation if any of the conditions  giving the standard, single-field, slow-roll is violated. These are: {\it a)} only one field is responsible for generating the primordial perturbations {\it b)} canonical  kinetic energy of the field {\it c)} slow roll {\it d)}  the quantum field was in the adiabatic (Bunch-Davies) vacuum. In particular the violation of  each of these conditions leaves its ``signature" on the statistical properties of the initial perturbations (e.g.,\cite{linde/mukhanov:1997,lyth/ungarelli/wands:2003, babich/creminelli/zaldarriaga:2004,chen/etal:2007,holman/tolley:2008,chen/easther/lim:2007,langlois/etal:2008, komatsuwhitepaper} and references therein). 

Deviations from Gaussianity can be characterized by the dependence of the bispectrum signal\footnote
{There are some cases where 
the trispectrum may be important (when, for example, 
the bispectrum is zero) but, in general, one expects the 
trispectrum contribution to be sub-dominant compared 
to the bispectrum one.}  on the  configuration (or {\it shape}) of the  three $k$-vectors in its argument.  For example {\it  local} type of non-Gaussianity yields a bispectrum that is dominated by the {\it squeezed} configurations (where $k_1\ll k_2\simeq k_3$) and is generated by models of inflation involving multiple fields\footnote{This is also the non-Gaussianity of standard slow-roll inflation, but, as mentioned above, in this case the amplitude is unmeasurably small, but see e.g., \cite{VM09}.}.
This shape is in general associated with models of inflation  where non-Gaussianity is created by non-linearities developed outside the horizon \cite{babich/creminelli/zaldarriaga:2004}.
When non-Gaussianity is created at horizon crossing during inflation, the primordial bispectrum is dominated by {\it equilateral} configurations ($k_1\simeq k_2 \simeq k_3$). Non-canonical kinetic terms will also yield  non-Gaussianities of this shape. 

The non-Gaussianities produced by the most general single-field inflation models with approximate shift symmetry have shapes that vary from being peaked on equilateral configurations to being peaked on enfolded ($k_1\simeq k_2\simeq k_3/2$) configurations \cite{senatore/smith/zaldarriaga:2009}. These types of non-Gaussianities, have been shown \cite{senatore/smith/zaldarriaga:2009} to be generically well described by  a linear combination of two  bispectrum shapes, one where the bispectrum signal peaks on equilateral configurations and another one,  orthogonal to it, called  {\it orthogonal} shape.   

Finally, modifications of the initial-state of the inflaton field, leave their signatures in a bispectrum dominated by {\it  flattened} or {\it enfolded} configurations \cite{chen/etal:2007, meerburg}.

The standard observables to constrain primordial non-Gaussianity are the cosmic microwave background (CMB) and large-scale structure. Both CMB and large-scale structure can measure, in principle,  the bispectrum shape dependence and thus discriminate the shape of non-Gaussianity. The large-scale structure bispectrum however, includes a large contribution form non-linear gravitational evolution and biasing; compared with this contribution, the primordial signal ``redshifts away" during the Universe's evolution \cite{VWHK00,V01,VJKM01}. On the other hand large-scale structure can probe scales non easily accessible from the CMB (e.g., see discussion in \cite{Loverdeetal07, Sefusatti}) and offers other powerful probes. One technique  is based on the abundance of rare events as they trace the tails of the underlying distribution \cite{MVJ00} and has received renewed attention e.g.,\cite{Loverdeetal07,Grossietal09,Pillepich,KJV09,JV09,Desjacques,verdereview} and references therein). Recently, the effect of non-Gaussian halo bias \cite{DDHS08,MV08} has been demonstrated to be an extremely promising avenue both for present \cite{afshordi/tolley:2008,slosar/etal:2008,xiaNVSS10} and future \cite{CVM08,CMV10} data.  This approach  uses the fact that the power spectrum of density extrema (where galaxies are formed) on large scales increases (decreases) for a positive (negative) $\fnl$. 
The signal, for these  two techniques  is given by an integral over the bispectrum. The shape-dependence is thus indirect and   suppressed, but signatures still remain \cite{KJV09,VM09}. 

While  techniques have been explored and developed to create non-Gaussian CMB maps with given bispectrum  and power spectrum for specific and  generic shapes of non-Gaussianities \cite{smith/zaldarriaga:2006, fergusson/liguori/shellard:2010}, to our knowledge,  cosmological  N-body simulations   with non-Gaussian initial conditions specified by a bispectrum have been  so far run only for local type of non-Gaussianity\footnote{Simulation were run e.g.  for $\chi^2$ initial conditions \cite{Robinson,Scoccimarro}, but  here we  are concerned with inflation-motivated non-Gaussianities classified by their bispectrum shape.}.  

Given the importance  N-body simulations have  in modeling both the non-linear physics and observational effects that play such a crucial role in interpreting large-scale structure data, here we demonstrate how to set up and run  N-body simulations with  non-Gaussian initial conditions specified by a generic bispectrum.
  
This paper is organized as follows.
In Sec. \ref{create} we outline how to create a three-dimensional non-Gaussian field with a given bispectrum. In Sec. \ref{implementation} we explicitly work out the expressions to use in the four most popular non-Gaussian shapes  discussed above and  detail the steps for implementation. We test the non-Gaussian initial conditions in Sec.~\ref{IC}. In Sec.~\ref{simulations} we describe our simulations and use the local case as a benchmark for our implementation. In Sec.~\ref{results} we present our results. We conclude in Sec.~\ref{conclusions}.

\section{Creating a 3D non-Gaussian field with a given bispectrum}
\label{create}
The argument of \cite{smith/zaldarriaga:2006} (see also \cite{fergusson/liguori/shellard:2010}) which applies to a two-dimensional field expanded in spherical harmonics,  can be generalized to apply  to a three-dimensional field transformed to Fourier space.  Pioneering work can be found in \cite{vio1, vio2}, here we follow a different route.
Let's start from 
\begin{equation} 
\label{eq:bispectrum0}
\langle \Phi_{k_1} \Phi_{k_2} \Phi_{k_3}\rangle=(2\pi)^3 \delta^D({\bf k}_1+{\bf k}_2+{\bf k}_3) B(k_1, k_2, k_3)\,.
\end{equation}
Here $B$ is intended  to be the bispectrum of the $\Phi$ field.  In this convention $\Phi$ is the Bardeen potential (i.e $-1 \times $ the standard gravitational potential, on sub-horizon scales). This equation  refers to some time deep in the matter-dominated era i.e., we use the $f_{\rm NL}$ CMB convention.

We can decompose the $\Phi$ field in a Gaussian and non-Gaussian parts. In Fourier space this reads
\begin{equation}
\Phi_{\bf k}=\Phi_{\bf k}^{G}+\Phi_{\bf k}^{NG}\,,
\label{eq:phgauss+ng}
\end{equation}
 and therefore 
 \begin{equation}
 \langle \Phi^G_{k_1} \Phi^G_{k_2} \Phi^{NG}_{k_3}\rangle=\frac{1}{3}(2\pi)^3  B(k_1, k_2, k_3)\delta^D({\bf k}_1+{\bf k}_2+{\bf k}_3)\,.
 \end{equation}
If we define 
\begin{eqnarray}
\label{eq:ansatz}
\!\!\!\Phi^{NG}_{\bf k}\!&=&\!\frac{1}{6(2\pi)^3}\!\!\int\! d^3k_2 d^3 k_3 B(k, k_2, k_3)\delta^D({\bf k}+{\bf k}_2+{\bf k}_3)\frac{\Phi^{*G}_{\bf k_2}\Phi^{*G}_{\bf k_3}}{P(k_2)P(k_3)}  \nonumber  \\
&=& \!\frac{1}{6(2\pi)^3}\!\!\int\! d^3k_2 B(k, k_2, |{\bf k}+{\bf k}_2|)\frac{\Phi^{*G}_{\bf k_2}}{P(k_2)}
\frac{\Phi^{G}_{{\bf k}+{\bf k}_2}}{P( |{\bf k}+{\bf k}_2|)}  
\end{eqnarray}
with $\Phi^*_{\bf k}$  the complex conjugate of $\Phi_{\bf k}$, 
and use it in Eq.~(\ref{eq:phgauss+ng}), we recover the identity of Eq.~(\ref{eq:bispectrum0}).

It is important to bear in mind  (as already pointed out in \cite{fergusson/liguori/shellard:2010})  that this procedure is not unique: in other words there may be more than one ---non-equivalent--- expression for $\Phi_k^{NG}$ yielding the same bispectrum. It is instructive to use the local non-Gaussian case as a worked example.
The  local case  is usually defined in real space  as \cite{Salopekbond90,VWHK00,KS01}:
\begin{equation}
\label{eq:local}
\Phi(\vx)=\Phi^G(\vx)+\fnl(\Phi^G(\vx)^2-\langle \Phi^G(\vx)^2\rangle)
\end{equation}
where $\Phi^G$ denotes a Gaussian random field. In Fourier space this becomes:
\begin{equation}
\Phi_\vk=\Phi^G_\vk+\fnl\frac{1}{(2\pi)^3}\int d^3k' \Phi^{*G}_{\vk'}\Phi^G_{\vk+\vk'}
\label{eq:localinfourier}
\end{equation}
where the constant has been absorbed in the $k=0$ mode which is ignored.

The bispectrum for the local non-Gaussianity is:
\begin{equation}
B(k_1, k_2, k_3)=2 f_{\rm NL}^{\rm local}[P(k_1)P(k_2)+ P(k_2)P(k_3)+P(k_1)P(k_3)]\,.
\label{eq:bisplocal0}
\end{equation}
Note that Eq.~(\ref{eq:localinfourier}) is {\it not} equivalent to Eq.~(\ref{eq:ansatz})\&(\ref{eq:phgauss+ng}).    Both procedures yield $\Phi$ fields with the same bispectrum, but the $\Phi$ fields obtained  are different.
In fact it can be shown that Eq.~(\ref{eq:ansatz})\&(\ref{eq:phgauss+ng}) become equivalent to Eq.~(\ref{eq:localinfourier}) only if $B(k_1,k_2,k_3)\longrightarrow 6 \fnl P(k_2)P(k_3)$  rather than using Eq.~(\ref{eq:bisplocal0}).
This extra ``degree of freedom" was used in Ref~\cite{fergusson/liguori/shellard:2010} to produce non-Gaussian fields with a given bispectrum  and with a minimal non-Gaussian contribution to the power spectrum. 
We return to this point below.

\section{Special cases and initial conditions implementation}
\label{implementation}

Let's start from  Eq.~(\ref{eq:bispectrum0}). The bispectrum  depends on the {\it shape} of the triangle made by the three $k$ vectors in its argument. 
The detailed dependence of the bispectrum on the $k$ vectors (also called in brief bispectrum shape)  can help discriminate among different inflationary models. For example,  local shape non-Gaussianities were the first type to be considered \cite{Salopekbond90, VWHK00,KS01} 
and are a direct consequence of the non-linear relation between the inflaton fluctuations 
and the curvature perturbations that couple to matter and radiation. While single-field slow-roll inflation generates this shape of non-Gaussianity, it has been shown that  the amplitude is proportional to the square of the slow-roll parameter, which is very small \cite{ABMR,maldacena:2003}.  In contrast, large local non-Gaussianities can be  generated in  e.g.,  curvaton models \cite{lyth/ungarelli/wands:2003}, where the curvature perturbation can evolve outside the horizon or   inflationary models with multiple scalar fields.   Non-canonical kinetic terms 
and higher derivative contributions to the inflaton potential can produce significant levels 
of non-Gaussianity of the equilateral type if the speed of sound in these models is much 
smaller than the speed of light, which can be realized in certain brane inflation scenarios \cite{seery/lidsey:2005,khoury/piazza:2009,chen:2005, silverstein/tong:2004}. These two shapes arise due to the interactions of the  field driving inflation.
 On the other hand, modified initial state (i.e. not starting from a Bunch-Davies adiabatic vacuum) also introduce deviations from Gaussianity. These are characterized by a different shape \cite{chen/etal:2007,meerburg} which we call {\it enfolded}.
Finally \cite{senatore/smith/zaldarriaga:2009} introduced the orthogonal shape: the  space of non-Gaussianities produced by the most general single-field models, where the inflaton fluctuations have an approximate shift symmetry, is spanned by linear combinations of two independent shapes: the equilateral and the orthogonal. We shall show below that these  four bispectrum types can be obtained from a linear combination of just three ``kernels'', which we will call $F^{\rm local}$, $F^{\rm A}$ and $F^{\rm B}$, speeding up our calculations.

The expressions for the bispectra for the four most  popular shapes mentioned above  are reported below.

In the local case:
\begin{equation}
\label{eq:B_loc}
B(k_1, k_2, k_3)=2 f_{\rm NL}^{\rm local} F^{\rm local}(k_1,k_2,k_3)\equiv f^{\rm local}_{\rm NL}{\cal B}^{\rm local}
\end{equation}

 where 
 \begin{equation}
\label{eq:Flocal}
 F^{\rm local}(k_1, k_2, k_3)=P(k_1)P(k_2)+ P(k_2)P(k_3)+P(k_1)P(k_3)\,,
 \end{equation}
 and we have  used ${\cal B}$ to denote the bispectrum for $\fnl=1$, and the dependence on the three $k$ vectors is understood.
 
In the equilateral case:
\begin{equation}
B(k_1, k_2, k_3)=6 f_{\rm NL}^{\rm eq} F^{\rm eq}(k_1,k_2,k_3)\equiv f^{\rm eq}_{\rm NL}{\cal B}^{\rm eq}
\end{equation}

where
\begin{eqnarray}
\label{eq:Feq}
F^{\rm eq}&=& -P(k_1)P(k_2)+2 {\rm cyc.} -2[P(k_1)P(k_2)P(k_3)]^{2/3}\\ \nonumber
&+& P^{1/3}(k_1)P^{2/3}(k_2)P(k_3) +5 {\rm cyc.}\\ \nonumber
 &\equiv&-F^{\rm local} -2F^A +F^B
 \end{eqnarray}
In the last equality we have introduced
\begin{eqnarray}
\label{eq:FAB}
F^A(k_1, k_2, k_3)&=& [P(k_1)P(k_2)P(k_3)]^{2/3} \\ 
F^B(k_1, k_2, k_3)&=&\left\{[P(k_1)]^{1/3}[P(k_2)]^{2/3}P(k_3) +5 {\rm cyc.}\right\}
 \end{eqnarray}
  and, for simplifying the notation, we have avoided writing down  the explicit  dependence of $F$ on the three $k$ vectors.
  
 For the  template for factorized  enfolded \cite{meerburg}: 
  \begin{equation}
  B(k_1, k_2, k_3)=6 f_{\rm NL}^{\rm enfl} F^{\rm enfl}(k_1,k_2,k_3)\equiv f_{\rm NL}^{\rm enfl}{\cal B}^{\rm enfl}
  \end{equation}
  where 
  \begin{eqnarray}
  F^{\rm enfl}&=&F^{\rm local}+3 [P(k_1)P(k_2)P(k_3)]^{2/3}\\ \nonumber
  &-&\left\{[P(k_1)]^{1/3}[P(k_2)]^{2/3}P(k_3) +5 {\rm cyc.}\right\}\\ \nonumber
&=& F^{local}+3F^A-F^B
  \end{eqnarray}
  
  Note that this is not the true enfolded configuration, but it is a factorizable template that approximates it well.

Finally, for the orthogonal shape:
\begin{equation}
 B(k_1, k_2, k_3)=6 f_{\rm NL}^{\rm orth} F^{\rm orth}(k_1,k_2,k_3)\equiv f_{\rm NL}^{\rm orth} {\cal B}^{\rm orth}
\end{equation} 
where 
\begin{equation}
\label{eq:Fort}
F^{\rm orth}(k_1,k_2,k_3)= -3F^{\rm loc}-8F^A+3F^B\,.
\end{equation}

It is important to note that these four shapes are {\it not} independent. For example the orthogonal one can be obtained from a combination of the other two:
\begin{equation}
F^{\rm orth}=F^{\rm eq}-2F^{\rm enfl}\,\,; \,\,\, {\cal B}^{\rm orth}= {\cal B}^{\rm eq} -2 {\cal B}^{\rm enfl}
\end{equation}

The factorization  introduced here speeds up the implementation if the initial conditions for N-body simulations of these four shapes:  initial conditions can be produced directly in Fourier space and split in a Gaussian realization $\Phi^G_{\vk}$ and three different non-Gaussian realizations $\Phi^{NG}_{\vk}$, one for the local case, one corresponding to $F^A$ and one to $F^B$.   From this one can build initial conditions for different shapes and different $\fnl$.

In the following we give some details on how we generate the initial particle distribution. We start off from the publicly available code by Sirko \cite{sirko2005} for Gaussian initial conditions and modify it appropriately. The difference between setting up Gaussian and non-Gaussian initial conditions is the extra non-Gaussian contribution, $\Phi^{NG}_{\vk}$, to the gravitational potential.  For a generic, non-factorizable bispectrum, we can  compute this field by disctretizing Eq.~(\ref{eq:ansatz}), i.e.,~for each grid point $\vk$ in Fourier space, we loop over all grid points $\bf k^\prime$:
\begin{equation}
\label{eq:discrete_ansatz}
\Phi^{NG}_{\vk}=\!\frac{1}{6}\!\!\sum_{\vk^{\prime}}\!  B(k, k^{\prime}, |{\bf k}+{\bf k}^{\prime}|)\frac{\Phi^{*G}_{\bf k^{\prime}}}{P(k^{\prime})}
\frac{\Phi^{G}_{{\bf k}+{\bf k}^{\prime}}}{P( |{\bf k}+{\bf k}^{\prime}|)}  \,,
\end{equation}
where $\Phi^{G}_{\vk}$ is a random realization of a Gaussian field with the power spectrum given by $P(k)\propto k^{n_s-4}$ and $n_s$ denotes the spectral index. If the number of grid points is given by $N_g^3$, the computational costs of the generation of such a generic non-Gaussian field $\Phi^{NG}_{\vk}$ scales as $N_g^6$. For a modest grid size of $256^3$, this results already in $~3 \times 10^{14}$ evaluations of the summand in Eq.~\ref{eq:discrete_ansatz}, which take about 1000 hours on a single core of a present-day CPU. For example, the computation for a $512^3$ grid would take approximately 10 days on 256 cores.

For factorizable shapes  the process can be greatly sped up. In fact note that in this case  Eq.~(\ref{eq:ansatz}) can be written as a convolution (or a sum of convolutions) of two auxiliary fields which evaluation can be swiftly done resorting to  Fourier transforms.
In fact if the bispectrum can be written  in a factorizable form as  $B(k_1,k_2,k_3)\equiv\sum _ib^i_1(k_1)b^i_2(k_2)b^i_3(k_3)$ then Eq.~(\ref{eq:ansatz}) becomes:

\begin{equation}
\!\!\!\Phi^{NG}_{\bf k}\!=\!\frac{1}{6}\!\sum_i b_1^i(k) \!\int\!\frac{d^3k_2}{(2\pi)^3}  G^i({\bf k}_2)Q^i({\bf k}+{\bf k}_2)
\end{equation}
where $G^i({\bf k})=b_2^i(k)\Phi^{*G}_{\bf k}/P(k)$ and $Q^i({\bf k})=b_3(k)\Phi^{G}_{{\bf k}+{\bf k}_2}/P( |{\bf k}+{\bf k}_2|)$.
The integral can then be quickly performed as a multiplication in real space (note the similarity with Eqs.~(\ref{eq:local}) and (\ref{eq:localinfourier}) where instead of  the $\Phi$ field we have two auxiliary fields $G$ and $Q$).

 While  for factorizable shapes the direct summation approach (Eq.~\ref{eq:discrete_ansatz}) and the convolution approach are mathematically equivalent in a practical implementation they may be affected by different numerical effects.  Here we explore both implementations for the local type. In the other cases  we implement the direct summation approach to explore whether this approach is viable and uncover possible bottlenecks or limitations.

Next, the linear density field $\delta_\vk$ at $z=0$ is derived from the potential $\Phi_\vk$ through the transfer function and the Poisson equation:
\begin{equation}
\label{eq:poisson}
\delta_{\vk}=\frac{2}{3}\frac{k^2T(k)D(z)}{\Omega_m H_0^2} \Phi_\vk\,,
\end{equation}
where $D(z)$ is the linear growth function normalized to $(1+z)$ in the matter-dominated era, and $T(k)$ is the transfer function obtained with CAMB \cite{CAMB} and normalized to unity on large scales. $\Omega_m$ is the present-day matter fraction and $H_0$ the Hubble constant.
The particles are then displaced from a regular grid according to the displacement field at the initial redshift, $z_i=49$, using the Zel'dovich Approximation\footnote{Since we will be interested  in the  non-Gaussian to Gaussian  ratio of our statistics, the detailed implementation of the initial displacement field (i.e.,  Zel'dovich  or   Second-Order Lagrangian perturbation theory) does not matter.}.

\section{Testing the initial conditions}
\label{IC}
In this section we analyze the quality of the non-Gaussian initial conditions. First we consider the 1-point function, i.e.~the probability distribution function of the density contrast $\delta$ at the position $x$. Especially, we compute the variance, $\langle \delta^2 \rangle$, and the skewness, $\langle \delta^3 \rangle$, as a function of the smoothing scale $R$ and compare the results with the analytic predictions. Furthermore, we calculate the power spectrum of the non-Gaussian initial conditions and demonstrate that the deviations from the Gaussian case are in most cases small.

\begin{figure}[htb]
\begin{center}
\includegraphics[angle=0,width=0.8\textwidth]{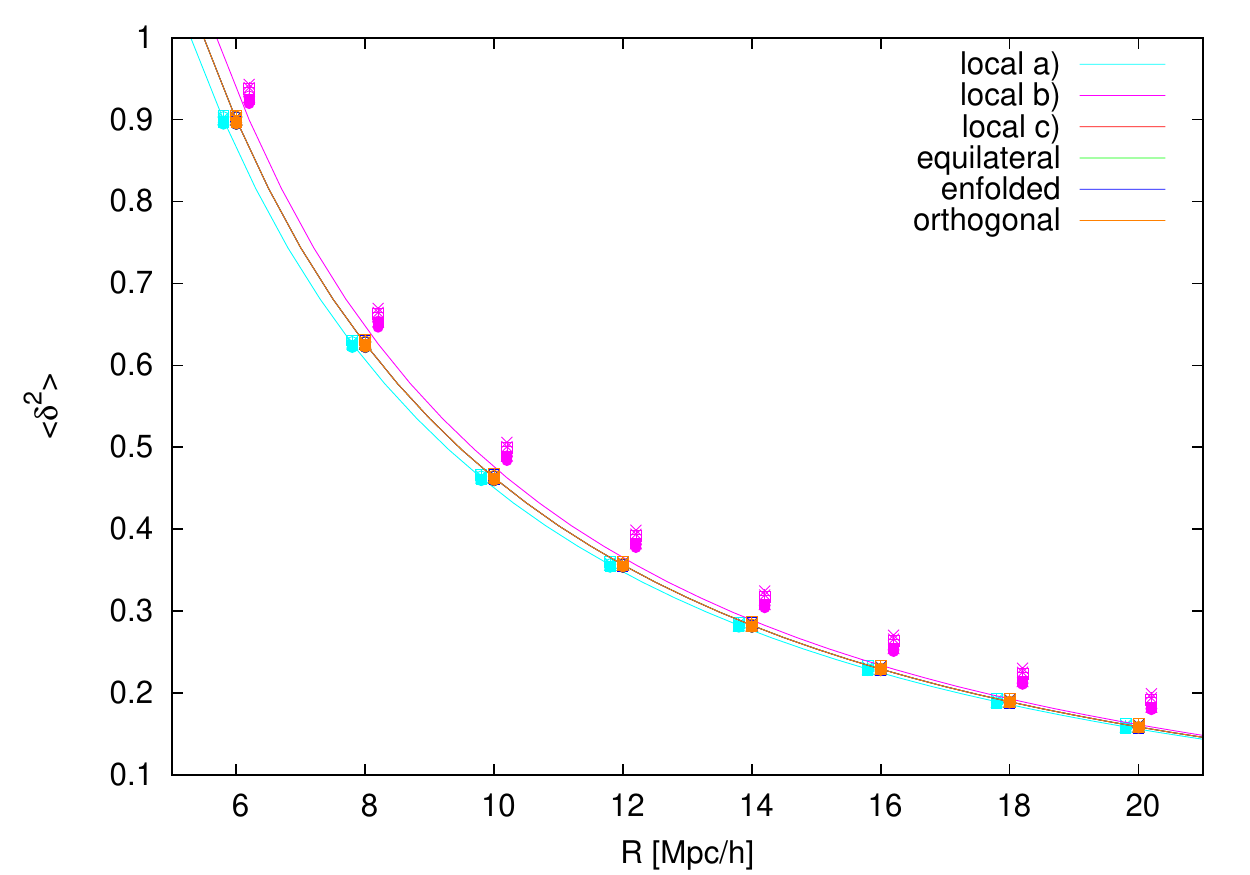}
\caption
{Variance of the linear density field at $z=0$ as a function of the smoothing radius $R$. The non-linearity parameter is $\fnl=100$. The symbols show the variance derived from the 8 different realizations of each type of non-Gaussianity. The lines depict the theoretical predictions. For clarity, the case ``local a)'' and ``local b)'' are slightly shifted horizontally. All other symbols and lines fall on top of each other.
}
\end{center}
\label{fig:variance}
\end{figure}

First we generate eight Gaussian realizations of the density field on a grid of size $256^3$ in a box with a side of $600\, {\rm Mpc}/h$. From these Gaussian realizations we produce for each of the previous mentioned types of non-Gaussianity, i.e.~local, equilateral, enfolded, and orthogonal, a non-Gaussian density field with an $\fnl$ of  $-500$, $-250$, $-100$, $100$, $250$, and $500$. 

For the local type, we compute the non-Gaussian contribution in three different ways: {\it a)} the traditional way by squaring the Gaussian density field in real space {\it b)} in Fourier space using our ansatz Eq.~(\ref{eq:discrete_ansatz}) with the bispectrum given by Eq.~(\ref{eq:bisplocal0}) {\it c)} using again Eq.~(\ref{eq:discrete_ansatz}) but with $B(k_1,k_2,k_3)\longrightarrow 6 \fnl P(k_2)P(k_3)$, this recovers the traditional method in real space except for aliasing effects introduced by the finite grid size.

\begin{figure}[htb]
\begin{center}
\includegraphics[angle=0,width=0.8\textwidth]{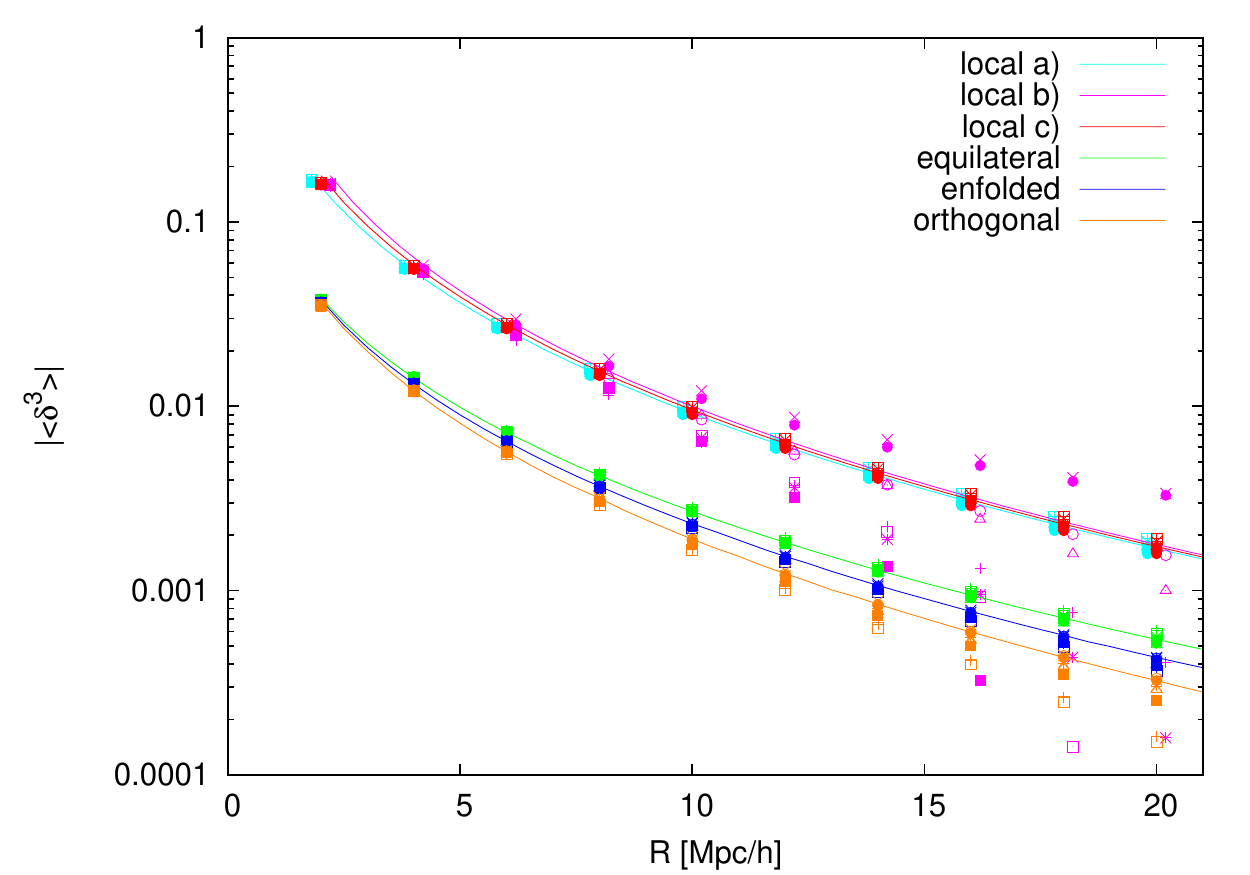}
\end{center}
\caption
{Skewness of the linear density field at $z=0$ for different types of non-Gaussianity with $\fnl=100$. Symbols show the skewness of the 8 different realizations. Note that the skewness of the Gaussian field due to the finite volume and grid size is subtracted from the measured skewness of the non-Gaussian field. Lines show the theoretical predictions.
Only the orthogonal case has a negative skewness for positive $\fnl$. For clarity, the case ``local a)'' and ``local b)'' are slightly shifted horizontally. 
}
\label{fig:skewness}
\end{figure}

In Fig.~\ref{fig:variance} and Fig.~\ref{fig:skewness} we show the variance and skewness of each realization (data points) for all types of non-Gaussianity with $\fnl=100$ and compare them to analytic predictions (solid lines). The analytic prediction for the skewness is obtained by integrating the bispectrum (see e.g., Eq.~(4.13) of Ref.~\cite{Loverdeetal07}).  We have truncated the integrals at the maximum and minimum $k$ sampled by the simulation box. The magnitude of the effect depends on  box size, scale  and type of non-Gaussianity. For our simulation settings, we find it always to be below 15\%. 

The moments of the density are computed from the linear density field smoothed with a top-hat filter of radius $R$. In case of the skewness, the $\langle \delta^3 \rangle$ of the Gaussian realization is subtracted from the total skewness in order to reduce the noise introduced by the finite volume and grid size. The variances agree well with the theory except for the type ``local b)''. The increased variance for this type is caused by the term $\langle \Phi^{NG}_\vk \Phi^{NG}_{-\vk}\rangle$, which, in this case,  is not negligible. This term gives rise to a $P^2(k)$ term multiplied with a divergent integral, which in our case of a finite box and grid is truncated. This is reminiscent of the discussion in sec. V of \cite{fergusson/liguori/shellard:2010};  for more details see  \ref{app:NG}. Apart from the ``local b)'' case, the values of the skewness obtained from the initial conditions also agree well with the predictions. Only for the orthogonal case small deviations at larger radii are visible.

\begin{figure}[htb]
\begin{center}
\includegraphics[angle=0,width=0.8\textwidth]{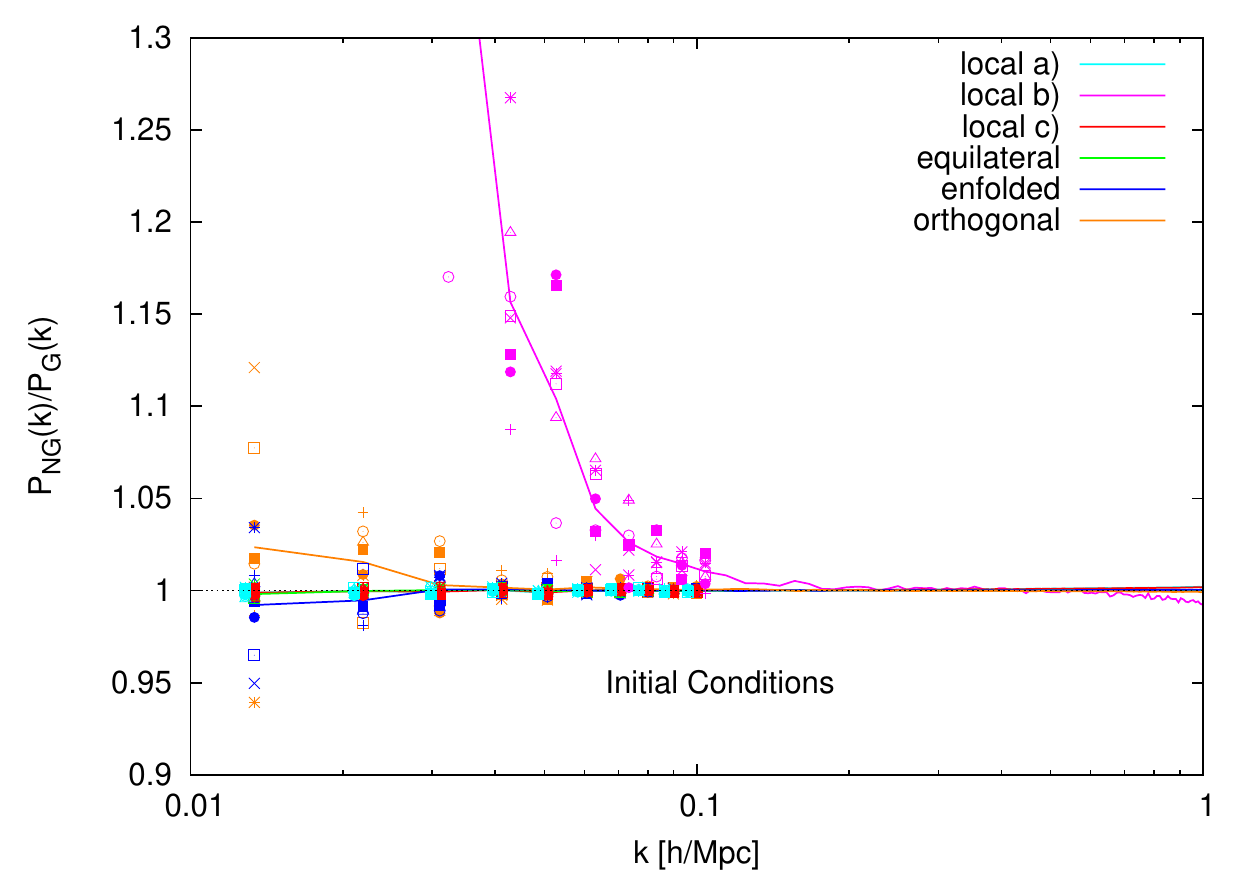}
\end{center}
\caption
{Ratio of the power spectra derived from the particle distributions of the non-Gaussian ($\fnl=100$) and Gaussian initial conditions. For $k<0.1\,h/{\rm Mpc}$, individual realizations are shown by symbols. The lines depict the arithmetic mean of the realizations. For clarity, the case ``local a)'' and ``local b)'' are slightly shifted horizontally.
}
\label{fig:Pk_IC}
\end{figure}

Next we consider the power spectrum of the initial particle distributions. We compute the power spectrum by first assigning the particle to a $512^3$ grid using the cloud-in-cell scheme. Then we perform a Fast Fourier Transform and average the $\vk$-modes over spherical shells with a thickness of $\Delta k = 0.01\,h/{\rm Mpc}$. The ratios of the power spectrum of the non-Gaussian and the Gaussian initial conditions are shown in Fig.~\ref{fig:Pk_IC}. The eight different realizations are depicted by the different symbols. For clarity, we show the individual realization only for $k\le 0.1\,h/{\rm Mpc}$. The solid line represents the mean of the realizations.
On large scales, the power spectrum of the ``local b)'' case deviates strongly from the Gaussian power spectrum. At the lowest wave number $k=0.01\,h/{\rm Mpc}$ the power spectrum is enhanced by almost an order of magnitude! This behavior explains the offset of the variance and can be traced back to the second-order term $\langle \Phi^{NG}_\vk \Phi^{NG}_{-\vk}\rangle$, which is further explored in the \ref{app:NG}. We observe a similar effect for the orthogonal shape, although the deviations are much smaller and a possible systematic shift, i.e.~enhancement or suppression of the power spectrum, is ---if at all--- at the few percent level. Nevertheless we discard the orthogonal shape from now on and refer to the \ref{app:NG} for more details on this issue. The case ``local a)'' and ``local c)'' agree perfectly with each other as it is expected from their mathematical equivalence. Both of these implementations of the local type and the equilateral case (almost not visible) do not show deviations from the Gaussian power spectrum. The power spectrum of the enfolded shape has very small deviations from the Gaussian power spectrum, the mean only deviates at the sub-percent level.
We also checked variance, skewness, and power spectrum for the other $\fnl$ values and found qualitatively similar results. The deviations from the Gaussian power spectrum scale roughly linear with $\fnl$, except for the ``local b'' case for which they scale as $\fnl^2$ (see \ref{app:NG} for details).

\section{Testing the simulation settings}
\label{simulations}
In this section we perform several tests to assure that the settings ---like force and mass resolution, box size, and initial redshift--- of our simulations are good enough to assure that systematic effects are smaller than the statistical errors. Note that we only consider ratios of non-Gaussian to Gaussian quantities in this paper. In particular, we are interested in ratios of the non-linear power spectrum and the halos mass function. Most of the systematic effects due to numerical limitations are expected to cancel out when considering ratios (see e.g.~\cite{McDonald2006}). In addition, since the non-Gaussian and Gaussian simulations are based on the same realization of the random density field, also the statistical error on the ratios is reduced vastly.

\begin{ctable}
  [caption={Settings of the N-body simulations.}, label={tab:sims}]
  {cccccccc}
	{\tnote{$\fnl=-500$, $-250$, $-100$, 0, 100, 250, 500}}
{\FL
  		  $N_{\rm part}$	& $L_{\rm box}$	& $m_{\rm part}$	& $l_{\rm soft}$ & $z_{\rm initial}$ & $\fnl$ & shape & \# sims \NN
	    	                & $({\rm Mpc}/h)$	& $(\MSUN/h )$& $({\rm kpc}/h)$	& 		&	&	& \ML
 $256^3$ & 600 & $10^{12}$ & 70 & 49 & $-500$ to 500\tmark & local c) & 8 \NN
 $256^3$ & 600 & $10^{12}$ & 70 & 49 & $-500$ to 500\tmark  & equilateral & 8 \NN
 $256^3$ & 600 & $10^{12}$ & 70 & 49 & $-500$ to 500\tmark & enfolded & 8 \NN
 $256^3$ & 600 & $10^{12}$ & 70 & 49 & $-500$ to 500\tmark  & orthogonal & 8 \ML
 $512^3$ & 600 & $\sim 10^{11}~~~$ & 35 & 49 & 0 and 100 & local a) & 1 \NN
 $512^3$ & 1200 & $10^{12}$ & 70 & 49 & 0 and 100 & local a) & 1 \NN
 $256^3$ & 600 & $10^{12}$ & 70 & 99 & 0 and 100 & local a) & 1 \NN
 $256^3$ & 600 & $10^{12}$ & 70 & 49 & 0 and 100 & local a) & 1 \LL
}

\end{ctable}

The standard settings of our main set of simulations can be found in the first half of Tab.~\ref{tab:sims}. The second half of the table describes the simulations which we use to investigate the numerical uncertainties of our main set of simulations. The generation of initial conditions done with the method ``local a)'' ---the traditional real-space implementation of local non-Gaussian initial conditions---  is by far not as computationally intensive as the method in Fourier space described in Sec. \ref{create} and \ref{implementation}. We exploit this fact and use the case ``local a)'' to perform larger N-body simulations to explore the numerical limitations of the smaller simulations of the main set. In addition, we use the case ``local a)'' as a benchmark for the implementation of local non-Gaussianity in Fourier space. 

All our N-body simulations are performed with the publicly available code Gadget-2 \cite{springel2005}. Our choice of cosmology is a flat $\Lambda$CDM cosmology, which is consistent with the seven-year WMAP results \cite{WMAP7}. In particular, we choose the following values: $\Omega_m=0.27$, $\Omega_b h^2 = 0.023$, $h=0.7$, $n_s=0.95$, and $\sigma_8=0.8$, where $\Omega_m$ is the matter density, $\Omega_b$ the baryon density, $h$ the Hubble parameter, $n_s$ the spectral index of the primordial power spectrum, and $\sigma_8$ the \textit{rms} of linear density fluctuations at $z=0$ in a sphere of $8\,{\rm Mpc}/h$.

\begin{figure}[htb]
\begin{center}
\includegraphics[angle=0,width=0.8\textwidth]{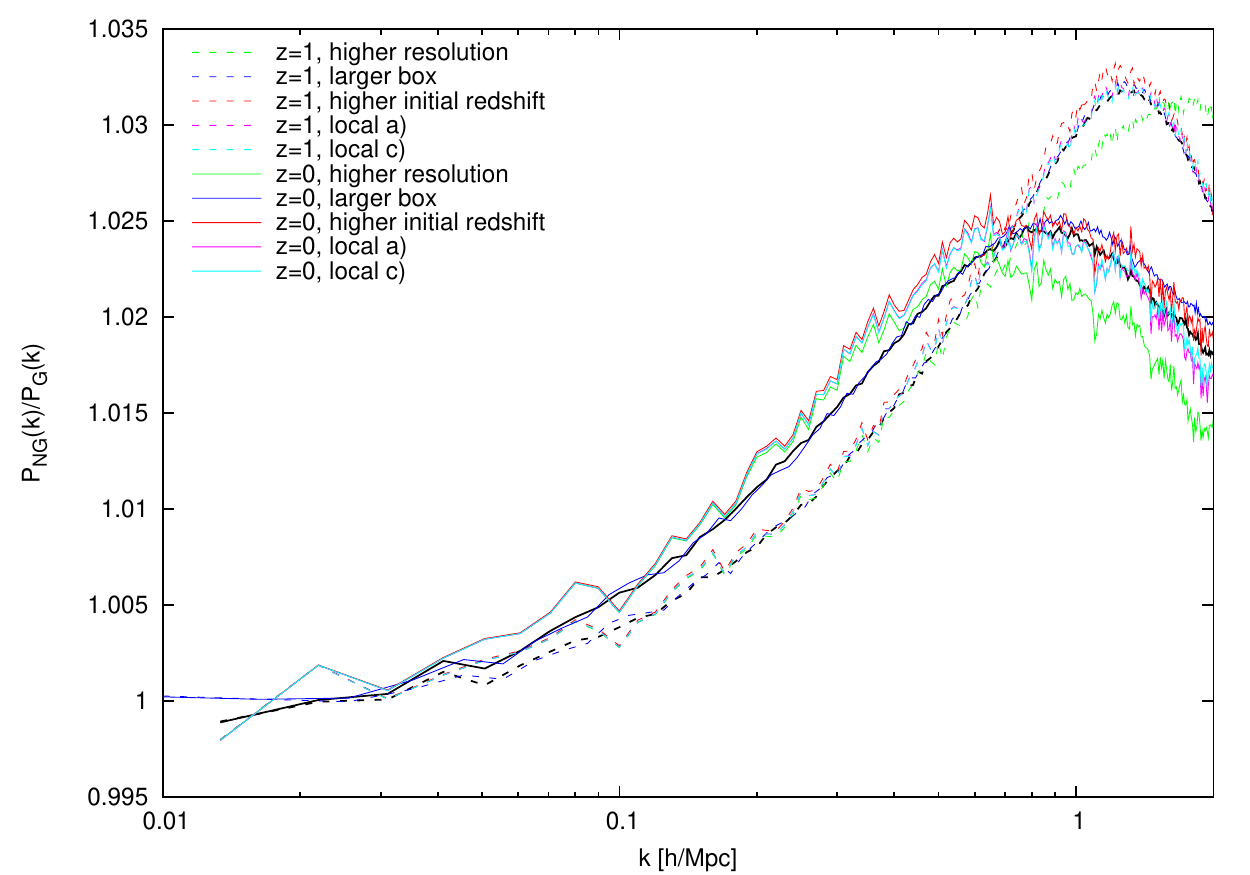}
\end{center}
\caption
{Ratio of the power spectrum of local non-Gaussian ($\fnl=100$) and Gaussian simulations at $z=1$ (dashed lines) and $z=0$ (solid lines) for different simulation settings (see Tab.~\ref{tab:sims}). The dashed and solid black lines show the mean ratio derived from the main set of our simulations of type ``local c)''.
}
\label{fig:Pk_benchmark}
\end{figure}

First, we consider the ratio of the non-linear power spectra derived from the local non-Gaussian simulations with $\fnl=100$ and Gaussian simulations at $z=1$ and $z=0$. The results are presented in Fig.~\ref{fig:Pk_benchmark}. The black dashed and solid lines depict the average of the eight realizations of the case ``local c)'' at redshift $z=1$ and $z=0$, respectively. The magenta and cyan lines show the two different implementations of the local type of non-Gaussianity for a single realization. The agreement is excellent and demonstrates the functionality of our method in Fourier space. The ratios derived from simulations of a larger box (1200 Mpc/$h$, blue lines) and from simulations with a higher starting redshift ($z_{i}=99$, red lines) are consistent with the ones derived from our main set of simulations. Only in the case of higher force and mass resolution, we find statistically significant deviations on smaller scales, i.e.~the ratio falls off for $k\gtrsim 0.7$ and $k\gtrsim 0.3$ at $z=1$ and $z=0$, respectively. The small reduction of the ratio is probably caused by the enhancement of non-linear power due to the better resolution.

\begin{figure}[htb]
\begin{center}
\includegraphics[angle=0,width=0.8\textwidth]{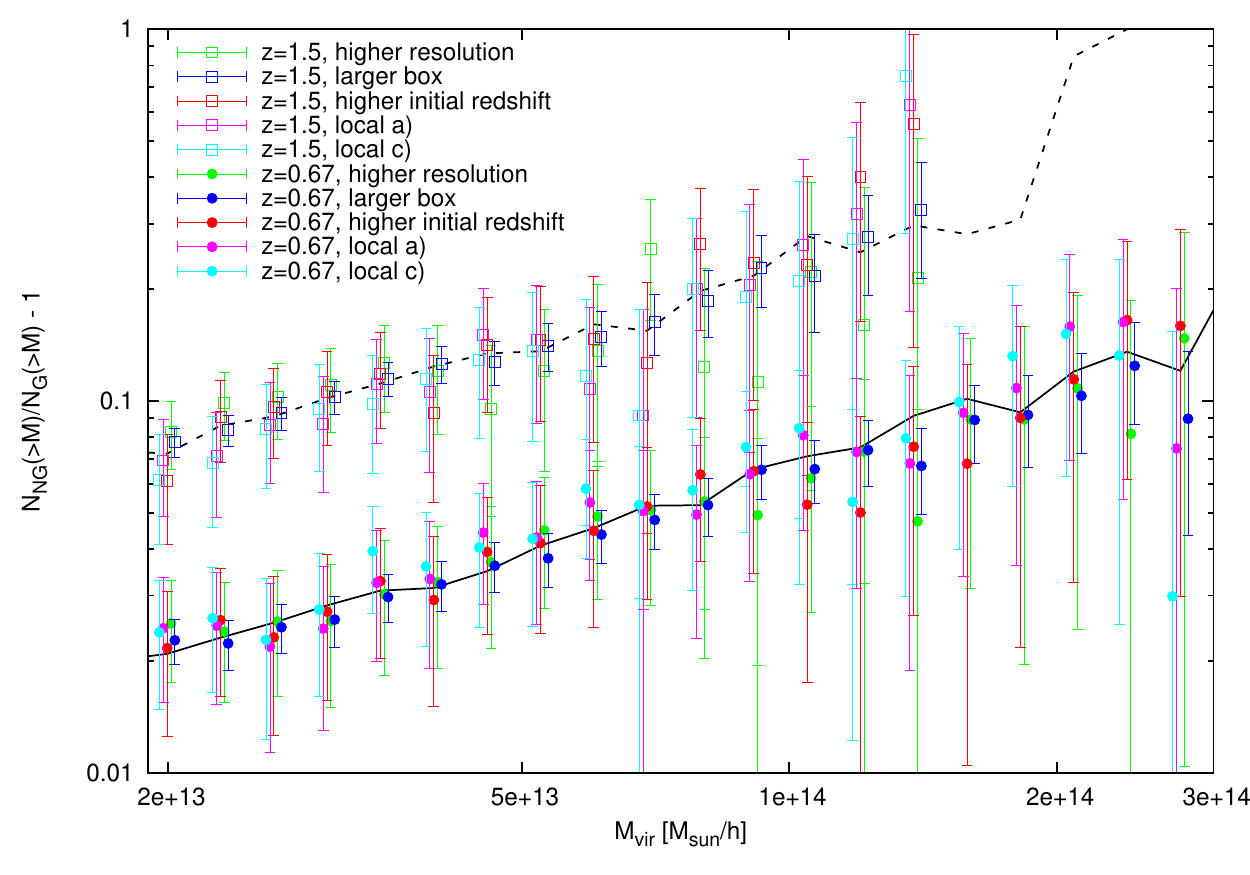}
\end{center}
\caption
{Fractional difference in the mass function of local non-Gaussian ($\fnl=100$) and Gaussian simulations at $z=0.67$ (filled circles) and $z=1.5$ (open squares) for different simulation settings (see text). Error bars represent Poisson errors. The dashed and solid black lines show the average of the fractional difference derived from the main set of our simulations of type ``local c)''.
}
\label{fig:mf_benchmark}
\end{figure}

The second quantity we are interested in is the ratio of the non-Gaussian and Gaussian halo mass function. To identify halos in the simulation data, we use the publicly available halo finder AHF \cite{Gill2004,knollmann2009}. AHF defines halos to be gravitationally bound objects which have a spherical overdensity (SO) given by the redshift dependent virial overdensity. The minimal number of particles in a halo is 20, thus in the main set of our simulations we find halos of mass $2\times 10^{13}\,\MSUN/h$ or higher. Although 20 particles are too few to reliably resolve all halos of the corresponding mass, we find  that the ratio of the mass functions is not affected by the low mass resolution. This is demonstrated in Fig.~\ref{fig:mf_benchmark}. The black dashed and solid lines show the averaged fractional difference in the non-Gaussian and Gaussian mass functions at redshift $z=1.5$ and $z=0.67$, respectively, derived from the eight simulations of the type ``local c)''. The results obtained from the simulations with an eight times higher mass resolution are depicted by the green symbols and are consistent with the lower resolution results. In addition, Fig.~\ref{fig:mf_benchmark} shows that the ratio is not biased by the finite-volume (blue symbols) nor the initial redshift (red symbols). The very good agreement between the two different ways of setting up the local case (cyan and magenta symbols) reassures us of the correct implementation of the method described in Sec. \ref{create} and \ref{implementation}.  

Overall the results of this section give us confidence that for our main set of simulations ---including the other types of non-Gaussianity--- systematic effects due to numerical limitations are within the statistical errors.

\section{Results}
\label{results}
Here, we present our findings derived from the non-Gaussian simulations of the local, equilateral, and enfolded type for $\fnl=-500$, $-250$, $-100$, $100$, $250$, and $500$. First we present the results for the non-linear matter power spectrum. Afterwards, we turn to the halos mass function. The investigation of  the non-Gaussian halo bias effect e.g., \cite{DDHS08, MV08} and the bispectrum is left to forthcoming  work \cite{CWinprep}.

\begin{figure}[htb]
\begin{center}
\includegraphics[angle=0,width=0.8\textwidth]{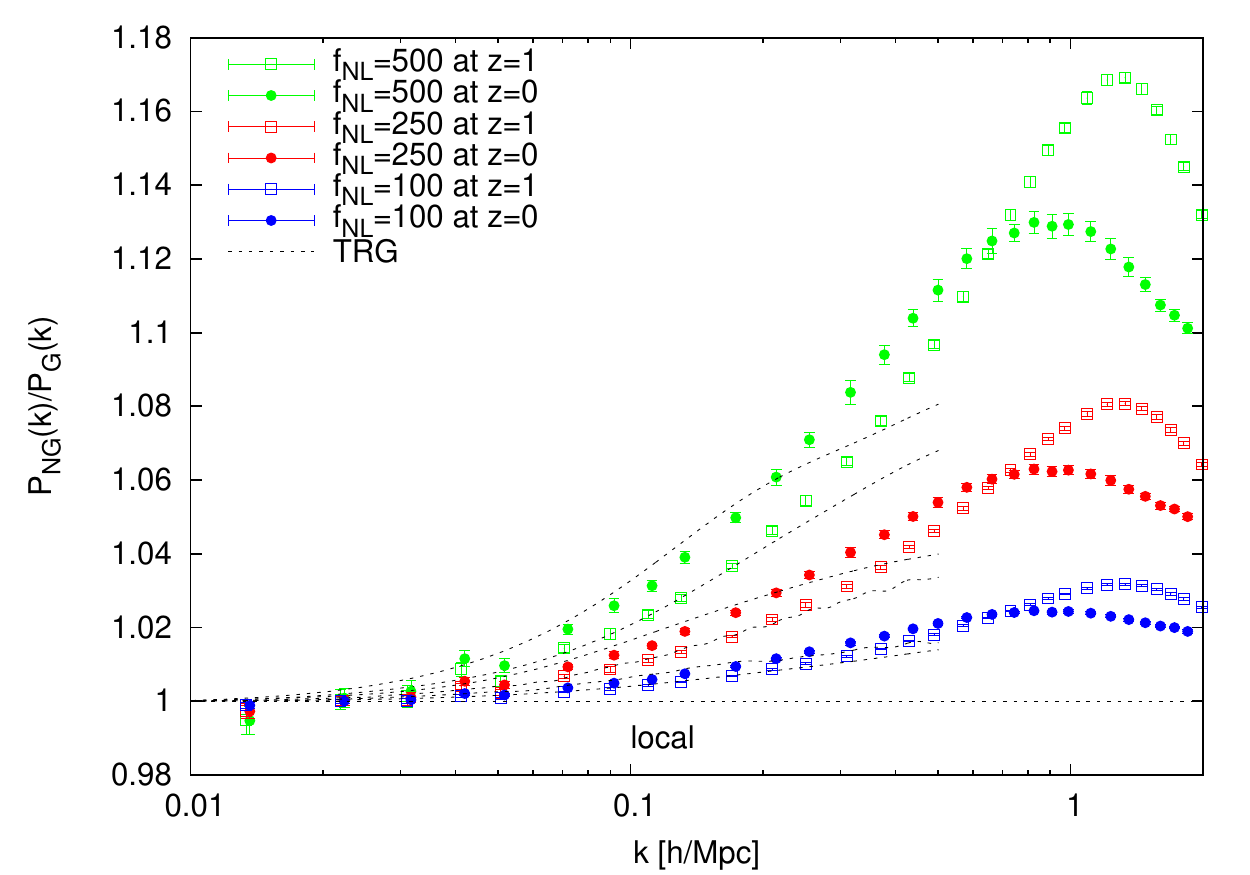}
\end{center}
\caption
{Ratio of the non-Gaussian and Gaussian power spectrum for the local type for different values of the non-linearity parameter $\fnl$ at $z=1$ (open squares) and $z=1$ (filled circles). The dotted black lines show the Time Renormalization Group (TRG) perturbation theory prediction of \cite{Bartolo2010}. Note we scaled the $\fnl=80$ data of \cite{Bartolo2010} linearly to $\fnl=100$. The error bars show the rms from the 8 realizations.}
\label{fig:Pk_loc}
\end{figure}

\begin{figure}[htb]
\begin{center}
\includegraphics[angle=0,width=0.8\textwidth]{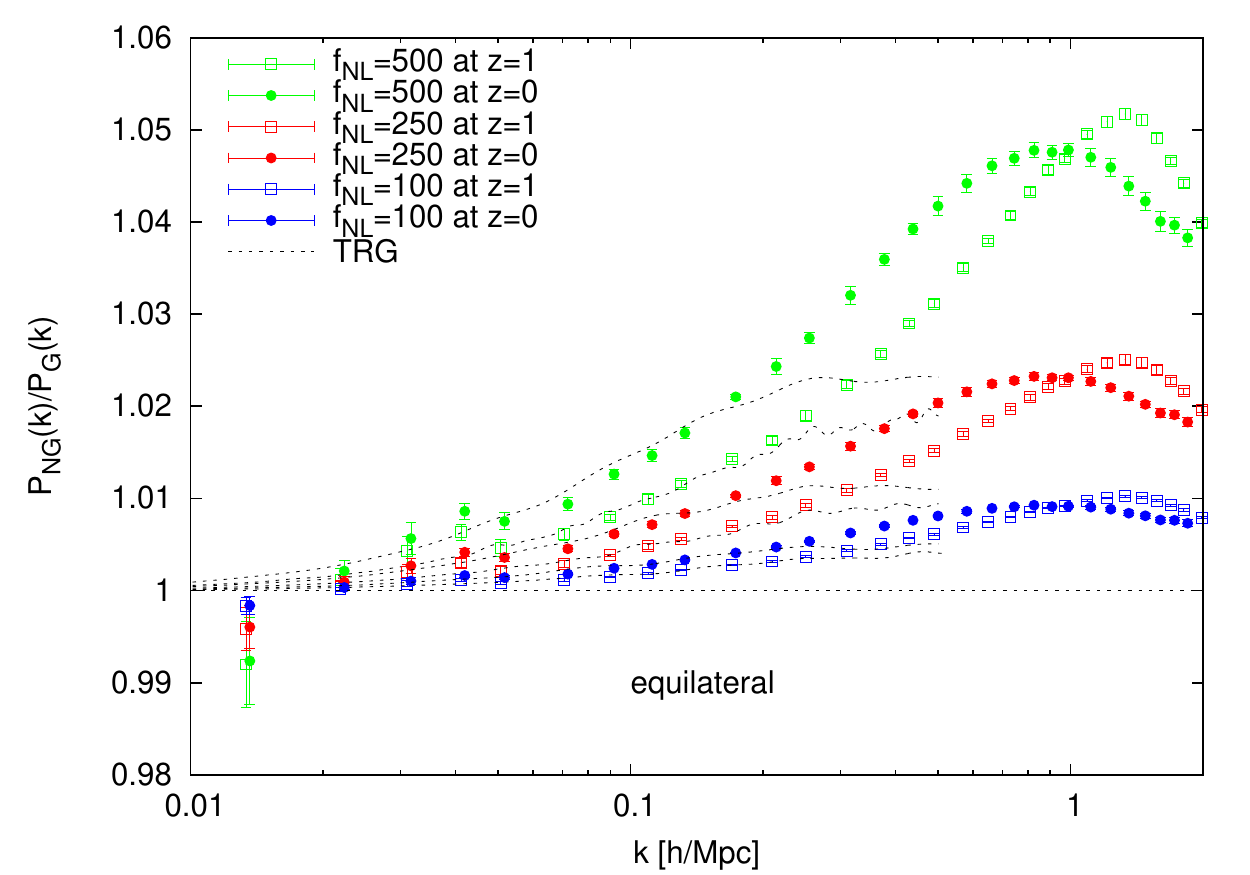}
\end{center}
\caption
{Same as in Fig.~\ref{fig:Pk_loc} but for the equilateral type of non-Gaussianity.}
\label{fig:Pk_eql}
\end{figure}

\begin{figure}[htb]
\begin{center}
\includegraphics[angle=0,width=0.8\textwidth]{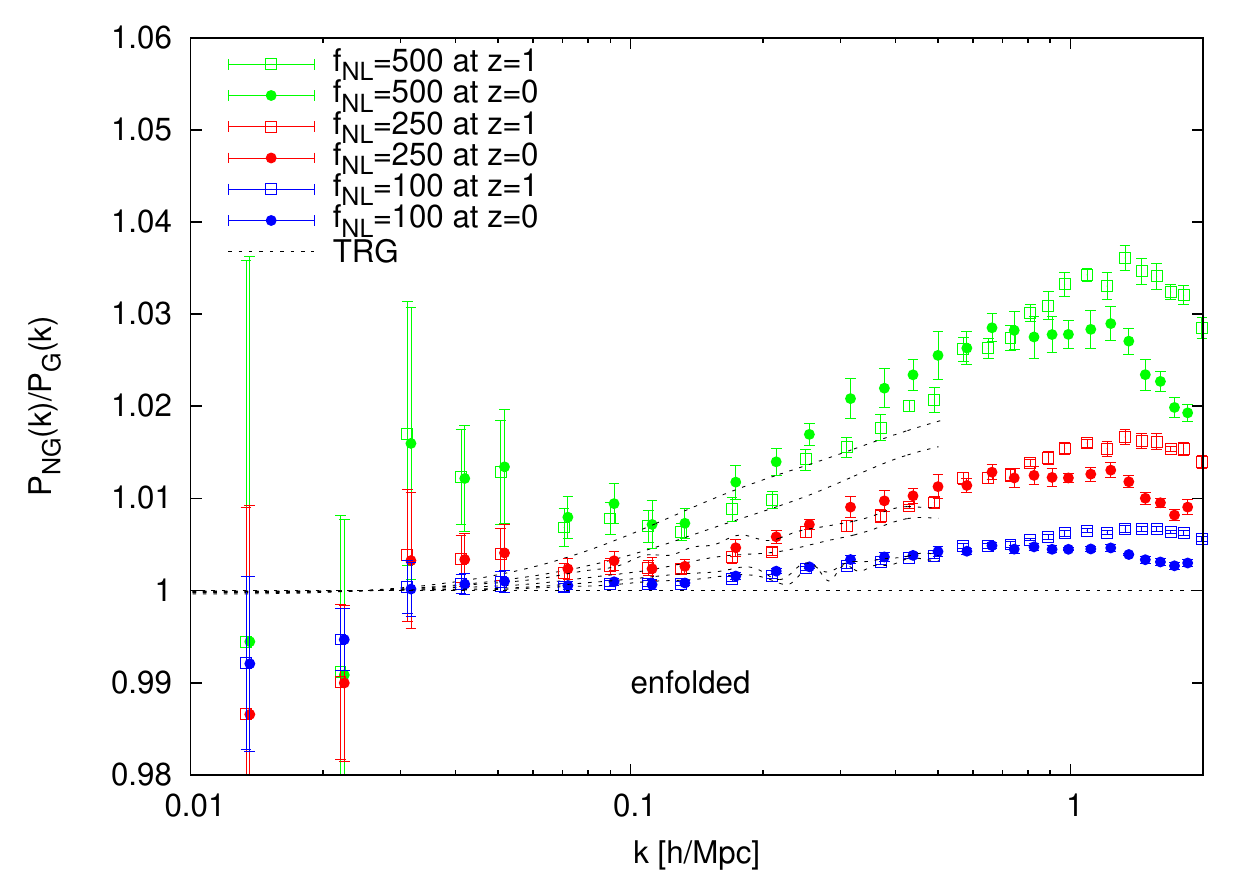}
\end{center}
\caption
{Same as in Fig.~\ref{fig:Pk_loc} but for the enfolded type of non-Gaussianity.}
\label{fig:Pk_enf}
\end{figure}

\subsection{Non-linear power spectrum}
\label{powerspectrum}
High precision in the theoretical prediction of the nonlinear matter power spectrum at $k \sim 1\,{\rm h/Mpc}$ and above  is needed to derive unbiased results from the data of upcoming weak-lensing surveys  e.g., \cite{Huterer/Takada2005}. Future Lyman-alpha forest surveys will also access these scales and test them with small statistical error-bars. Non-Gaussianities in the initial conditions alter the nonlinear power spectrum at the few percent level. In Fig.~\ref{fig:Pk_loc} (local), Fig.~\ref{fig:Pk_eql} (equilateral), and Fig.~\ref{fig:Pk_enf} (enfolded), we show the ratio of the non-Gaussian and Gaussian power spectrum for different type of non-Gaussianities with $\fnl=500$ (green), 250 (red), and 100 (blue) at redshift $z=1$ (open squares) and $z=0$ (filled circles). The black dashed lines correspond to the perturbation theory prediction using the time-renormalization group (TRG) approach \cite{Bartolo2010}. For $z=1$ and $k\lesssim 0.2\,h/{\rm Mpc}$, the TRG predictions agree very well with the results of the N-body simulations. At $z=0$, perturbation theory slightly overpredicts the effect on scales around $k\sim 0.1\,h/{\rm Mpc}$ before it breaks down on smaller scales ($k\lesssim 0.2\,h/{\rm Mpc}$).

Note that the maximum of the ratio is larger at higher redshifts and its positions is shifted to smaller scales. The shape, especially the peak location and peak height, is consistent with predictions of the halo model presented in \cite{fedeli2010}.

\begin{figure}[htb]
\begin{center}
\includegraphics[angle=0,width=0.49\textwidth]{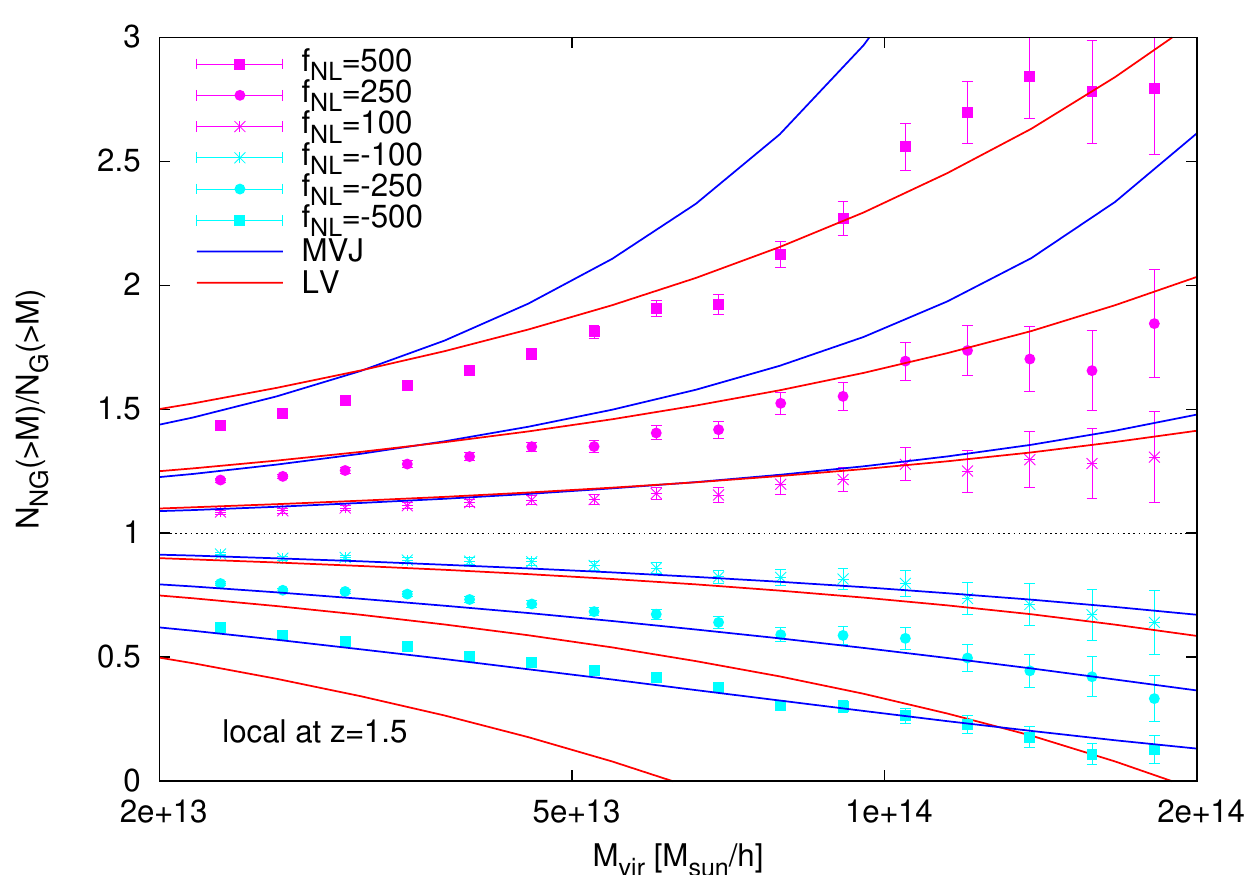}
\includegraphics[angle=0,width=0.49\textwidth]{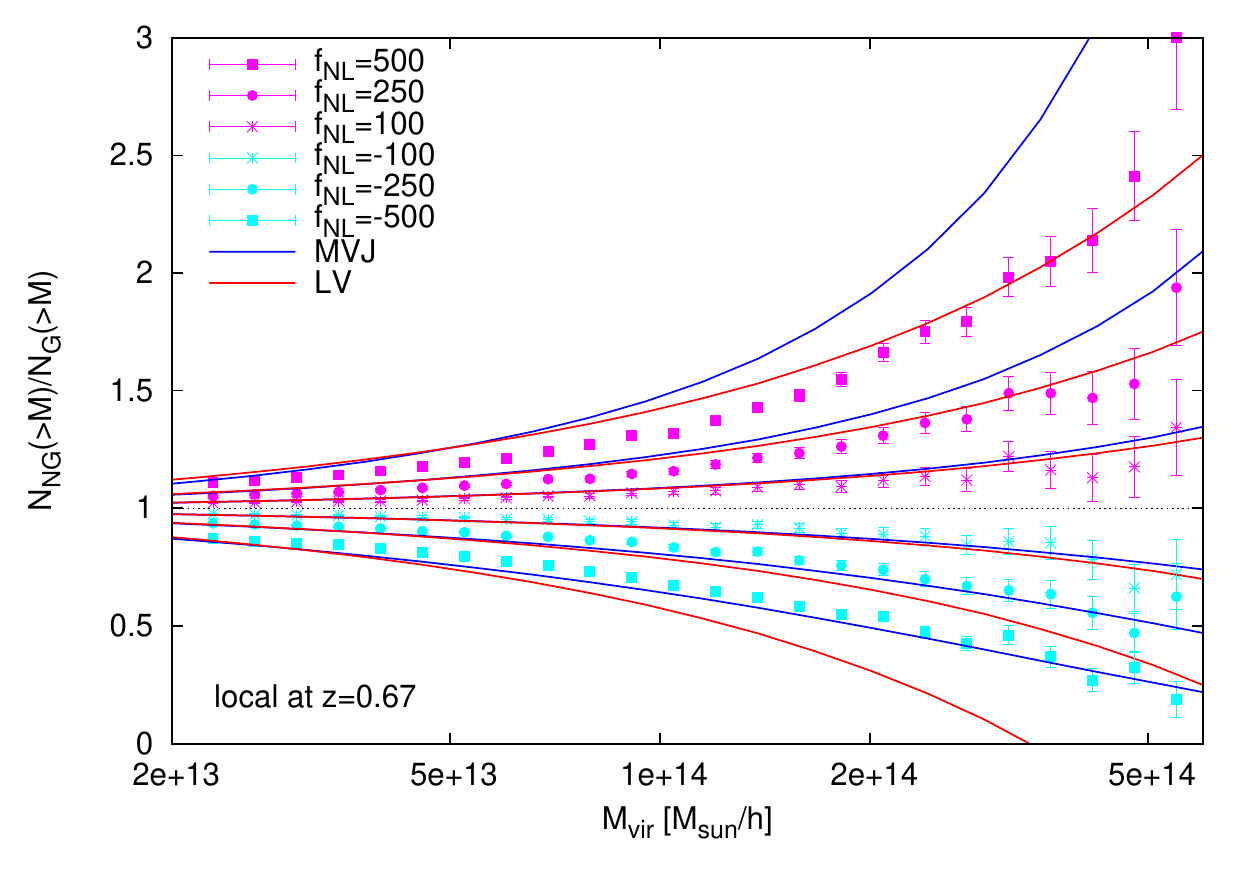}
\includegraphics[angle=0,width=0.49\textwidth]{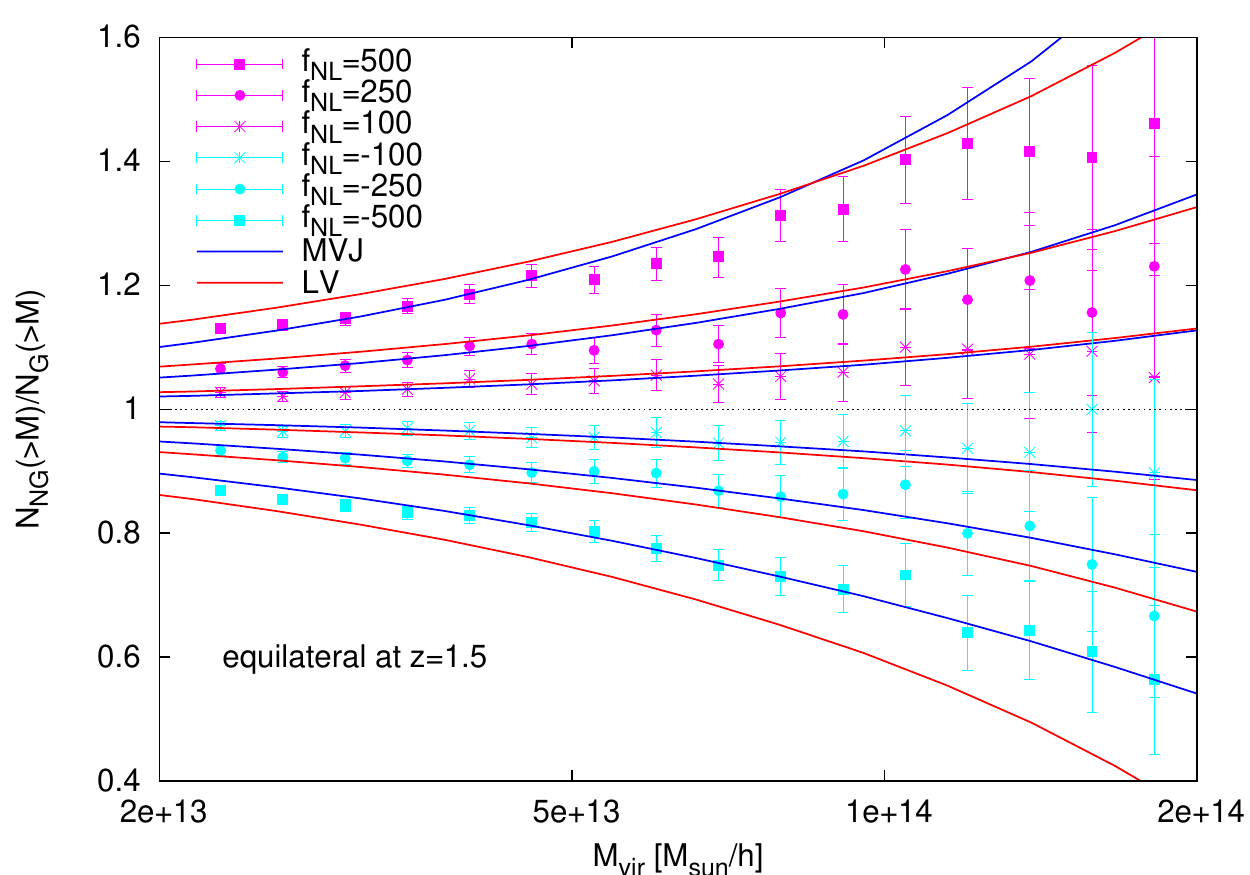}
\includegraphics[angle=0,width=0.49\textwidth]{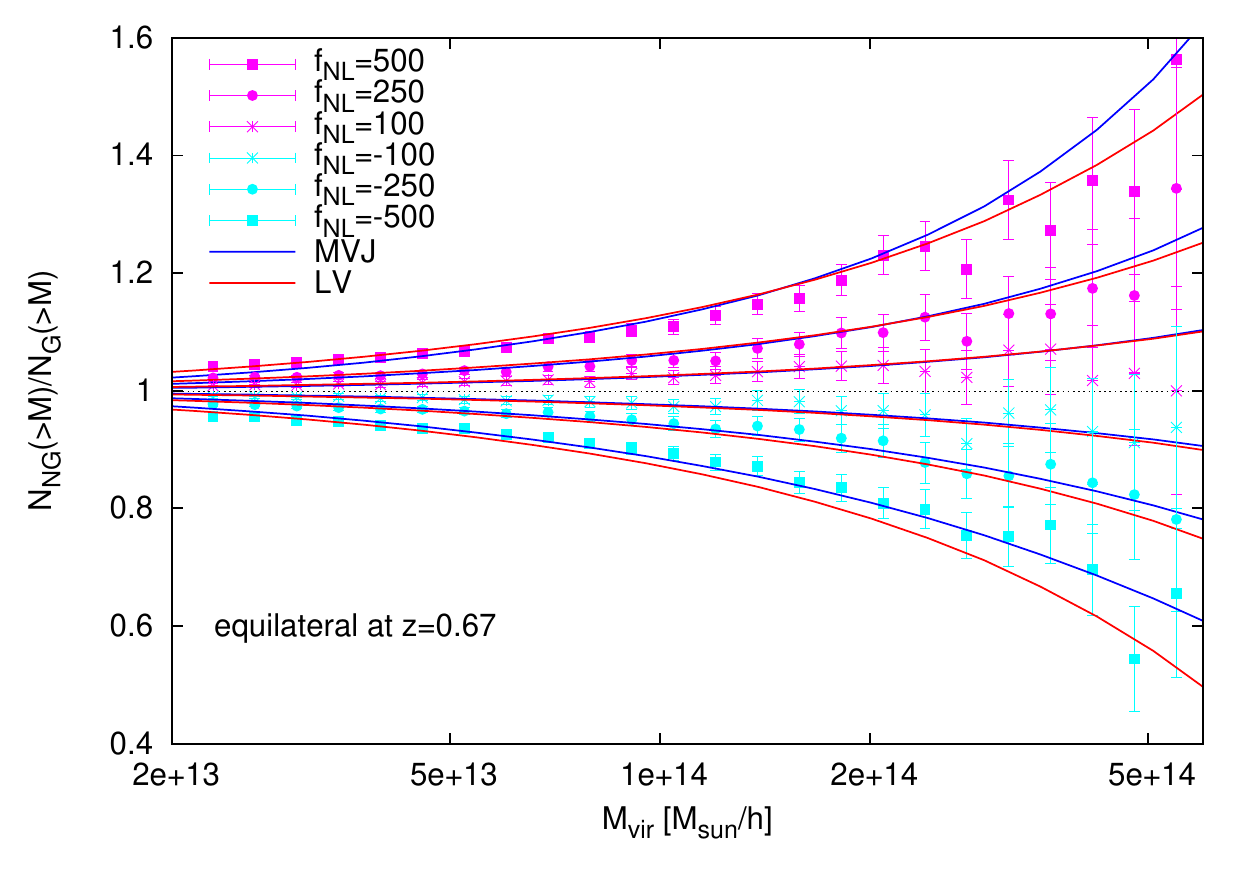}
\includegraphics[angle=0,width=0.49\textwidth]{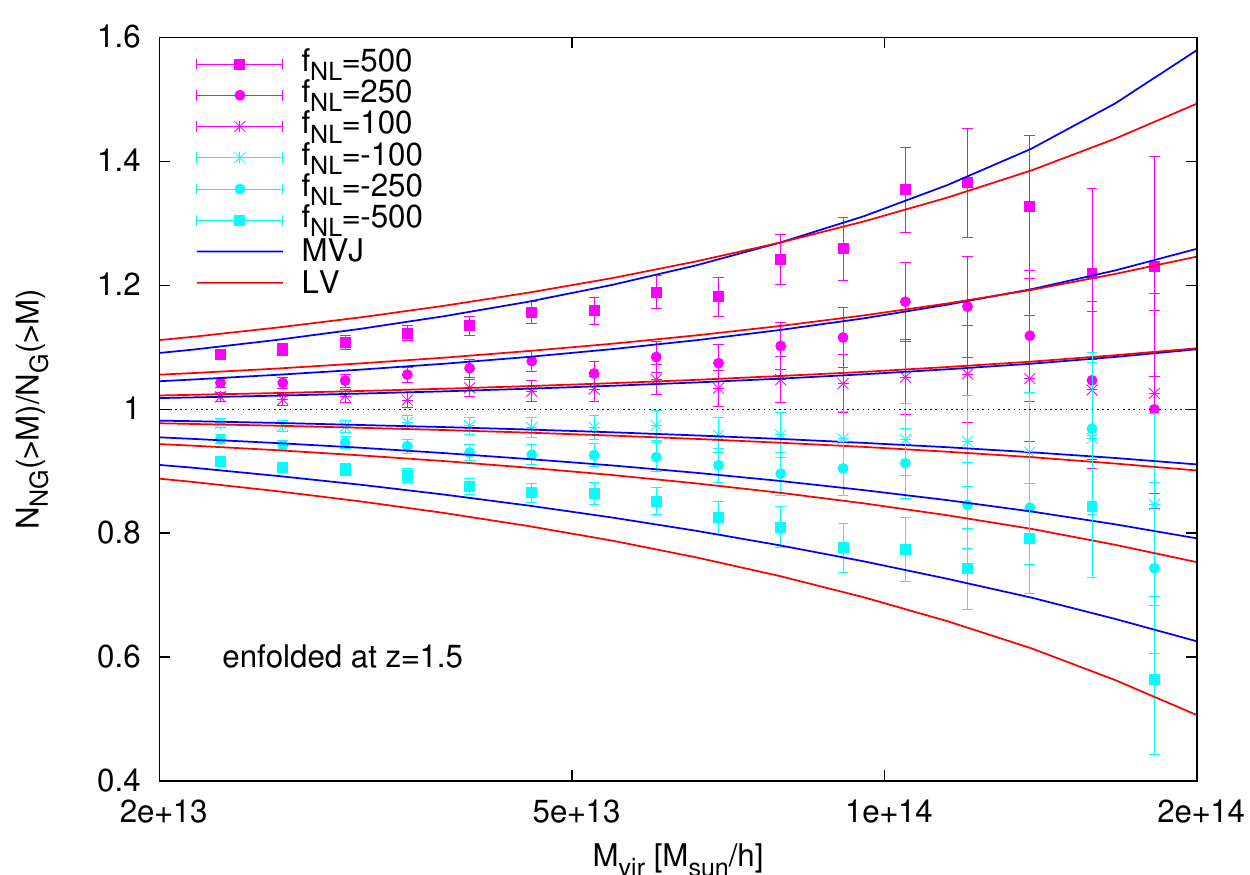}
\includegraphics[angle=0,width=0.49\textwidth]{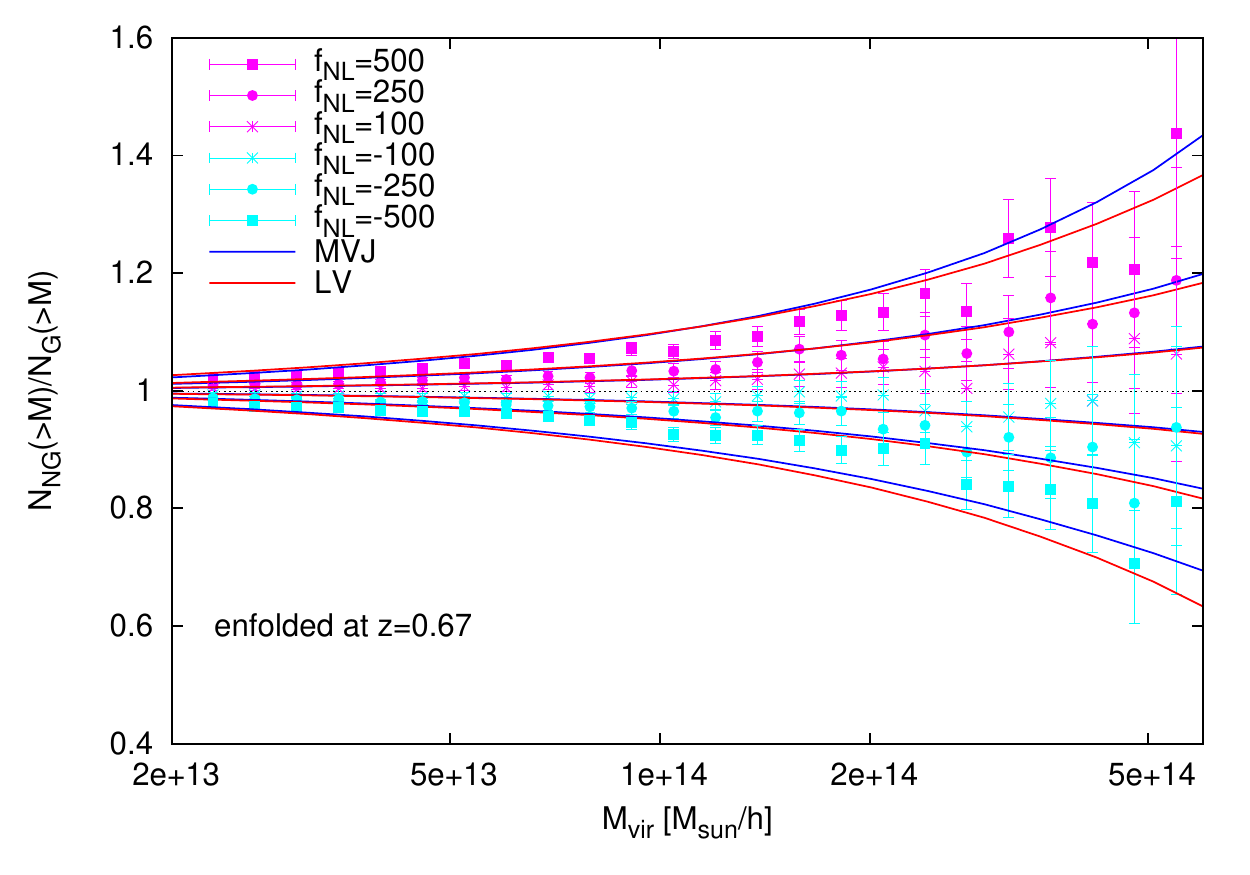}
\end{center}
\caption
{Ratio of the non-Gaussian and Gaussian \textit{cumulative} mass functions for the local, equilateral and enfolded shape at redshift $z=1.5$ and $z=0.67$ for different values of $\fnl$. The symbols depict the results derived from the simulations using 8 different realizations for each type of non-Gaussianity. For the error bars, we assume Poisson errors. The blue solid lines show the predictions of \cite{MVJ00} and the red lines represent the model of \cite{Loverdeetal07}.
}
\label{fig:MF_ratios}
\end{figure}

\subsection{Halo mass function}
\label{MF}
At the high-mass end, the halo mass function is very sensitive to a primordial skewness in the probability distribution function of the density field \cite{Lucchin/Matarrese1988}. Hence, galaxy cluster surveys offer in principle the possibility to probe primordial non-Gaussianities. In Fig.~\ref{fig:MF_ratios} we present the ratios of the {\it cumulative} mass functions derived from the non-Gaussian and Gaussian simulations.  We show the results for the local, equilateral, and enfolded shape with different $\fnl$ at redshift $z=1.5$ and $z=0.67$. In addition to the data points we plot the analytic predictions of \cite{MVJ00} (MVJ, blue lines) and \cite{Loverdeetal07} (LV, red lines). 
In order to convert the analytic ratio of the non-Gaussian and Gaussian mass functions, $r_{NG}(M)$, provided in \cite{MVJ00} and \cite{Loverdeetal07} into the corresponding ratio of the cumulative mass functions, $R_{NG}(>M)$, we use the fitting formula of \cite{Tinker2008} for the SO halo mass function, $n_{\rm Tinker}(M)$, i.e.
\begin{eqnarray}
R_{NG}(>M)=\frac{\int_M^\infty r_{NG}(\tilde M)n_{\rm Tinker}(\tilde M)d\tilde M}{\int_M^\infty n_{\rm Tinker}(\tilde M)d\tilde M}\,.
\end{eqnarray}
We checked that the integrated Tinker fit is a good fit to the cumulative mass function of the SO halos found in our Gaussian simulations.

Note that, compared  to \cite{Grossietal09}, we do not find that substituting the linear spherical collapse overdensity by $\delta_c \longrightarrow \sqrt{q}\delta_c$ (with $q=0.75$) improves the agreement between the N-body data and the predictions. This is in accordance with the findings of \cite{Desjacques2010a,Desjacques2010b}, who argue that the differences are due to the different halo identification algorithms. While in this paper we use halos defined by spherical overdensities, \cite{Grossietal09} applied a Friends-of-Friends halo finder to their simulations. 

At the high-mass end, we find that LV fits the data better for positive $\fnl$, whereas MVJ gives better predictions for negative $\fnl$. In general, for large positive $\fnl$ the theoretical predictions overestimate the increase of halo number density at the low-mass end. However, for all shapes, the qualitative behavior is captured well by the analytic models.
There is an indication that using spherical overdensities halos a value  $q\sim 0.9$ yield a slightly better fit to the simulations data. Given our simulations size, this  effect is however  not highly significant and only visible for high values of $f_{\rm NL}$. While the exact value of $q$ is an important theoretical issue, recall that a 10\% uncertainty in $q$ will propagate into a 10\% error on $f_{\rm NL}$.  For values of  $f_{\rm NL} \lesssim 10$, as expected from most models,   this  uncertainty is  at most comparable to the expected  statistical errors.

\section{Conclusions and discussion}
\label{conclusions}
We have addressed the issue of setting up  generic non-Gaussian initial conditions for N-body simulations.   In the era of precision cosmology, N-body simulations have become  a crucial tool to test,  develop and calibrate  any statistical analysis of large-scale structure surveys. While the  approach of constraining  primordial non-Gaussianity with large-scale structure and galaxy surveys  has recently received renewed attention, as far as we know, N-body simulations have been run only for the so-called local-type of non-Gaussianity. The shape of non-Gaussianity, however, is  crucial if one wants to use such a signal  as a window into the generation mechanism for primordial perturbations.

Building  on the expertise developed in the context of Cosmic Microwave Background non-Gaussianity, we have shown how to set up non-Gaussian initial conditions for any type of non-Gaussianity  specified by a primordial bispectrum. Given the current  cosmological constraints  all non-Gaussian fields we consider are given by the sum of a (dominant) Gaussian  component and a (subdominant) non-Gaussian one.

  The implementation is based on direct summation in Fourier space and is  significantly more computationally  intensive than the workhorse local case defined in real  space, as it scales as $N_g^6$ where $N_g^3$ is the number of grid points in the simulation box.  
We have investigated the numerical effects of such an approach  by comparing the results of the local-type case obtained with the two approaches.
For factorizable templates, the implementation can be sped up significantly: in fact the  operation can be rewritten as a convolution of two suitably defined auxiliary fields. This is  useful, however not the main goal of our investigation, as our aim was to develop an approach suitable for non-factorizable templates. This is relevant because for some physically motivated inflationary models the  existing factorizable templates may not be a good approximation of the effect e.g.,\cite{FergussonShallard,Chen, strings} and references therein.  

The procedure of generating a non-Gaussian field  with a given bispectrum (and a given power spectrum for the Gaussian component) is not univocal and care must be taken so that higher-order corrections do not leave a too large signature on the power spectrum especially on large scales. This is so far a limiting factor of our approach, as these components can be kept under control only in specific cases. More investigation is clearly needed, possibly along the lines proposed by \cite{fergusson/liguori/shellard:2010}. 
 
We have then explored  the effects of several popular forms of primordial non-Gaussianity on  the halo-mass function and the non-linear power spectrum deferring the analysis of other statistics such as the non-Gaussian halo bias to a forthcoming paper.  We confirm that the non-Gaussian correction to the halo mass function is determined by the primordial skewness and that a suitable combination  of the different analytical approximations proposed in the literature (depending on the regime of applicability) offer a  good fit to the simulations data. We also confirm that independently of the type of non-Gaussianity  spherical-overdensity halos  need a $q$ correction factor much closer to unity (if any correction at all)  than Friends-of-Friends halos. 
   
We believe that the methodology illustrated and developed here will be relevant for  the on-going and planned efforts of constraining  primordial non-Gaussianity  from large-scale structure  including  surveys of  galaxies, high-redshift clusters, Lyman-alpha forest and  weak lensing.

\section*{Acknowledgments}
CW is supported by  MICINN grant AYA2008-03531.
LV acknowledges support from FP7-PEOPLE-2007-4-3-IRG n.~202182 and FP7-IDEAS Phys.LSS 240117 and  MICINN grant AYA2008-03531. LV thanks Michele Liguori for stimulating discussions. LB acknowledges financial support from Spanish MEC and FEDER (EC) under grant FPA2008-02878 and Generalitat Valenciana under the grant
PROMETEO/2008/004. LB thanks the Institute of cosmological sciences
ICC-ICREA Barcelona for hospitality.

\appendix
\section{Non-Gaussian contributions to the initial power spectrum}
\label{app:NG}
As before we write the gravitational potential as the sum of a Gaussian and non-Gaussian field, $\Phi_\vk=\Phi^G_\vk+\Phi^{NG}_\vk$, where the non-Gaussian part is quadratic in $\Phi^G$. The power spectrum, defined by $\langle \Phi_\vk \Phi_{\bf q}\rangle = (2\pi)^3 \delta^D(\vk +{\bf q}) P(k)$, consists then of the following terms $\langle \Phi^G \Phi^G\rangle$, $\langle \Phi^G \Phi^{NG}\rangle \sim \fnl \langle \Phi^G \Phi^G \Phi^G \rangle$, and $\langle \Phi^{NG} \Phi^{NG}\rangle \sim \fnl^2 \langle \Phi^G \Phi^G \Phi^G \Phi^G \rangle$. The term $\langle \Phi^G \Phi^{NG}\rangle$ will vanish in theory as it involves an odd number of Gaussian fields. In practice, we have only a finite number of modes to perform the average. Hence, especially on larges scales where there are only a few modes in the box, this term is not exactly zero. We investigate this term further below.

\begin{figure}[htb]
\begin{center}
\includegraphics[angle=0,width=0.8\textwidth]{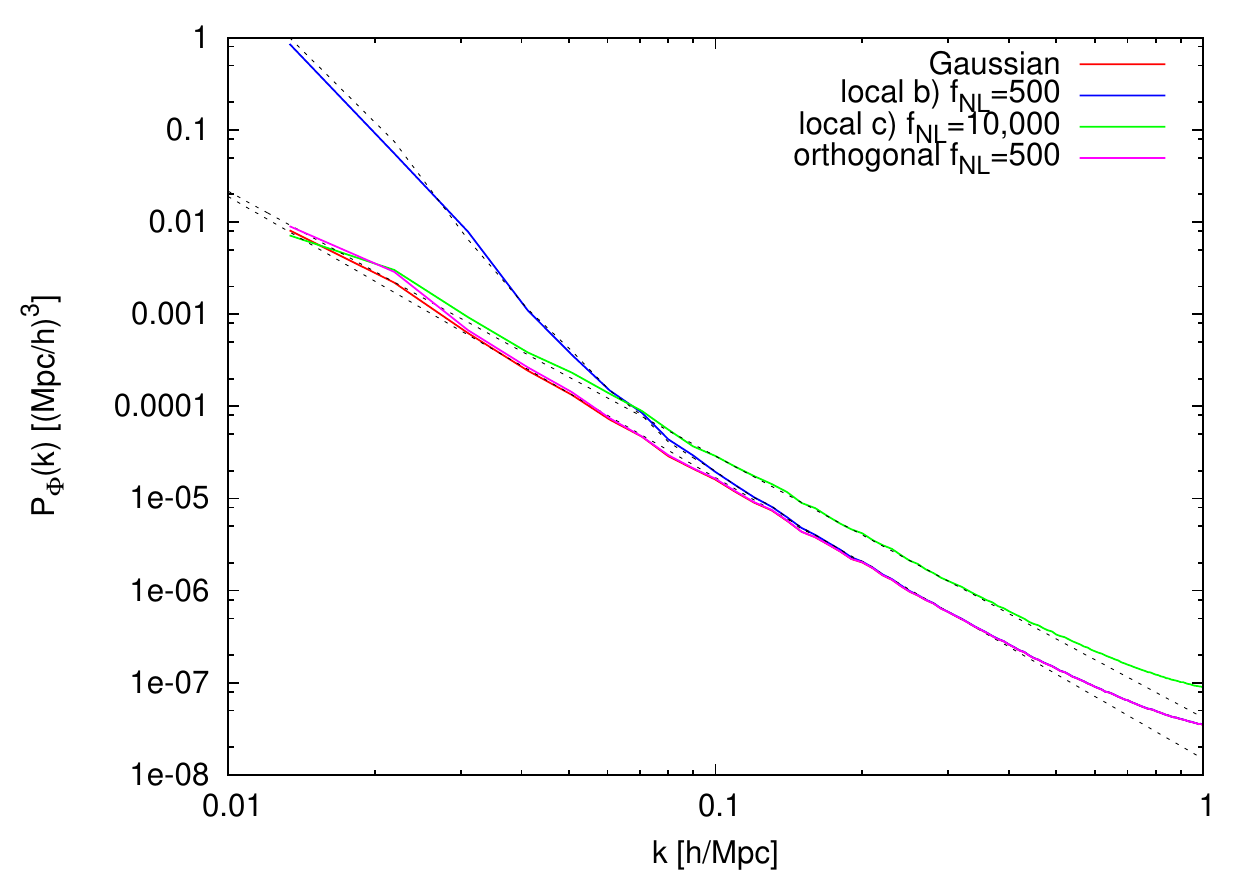}
\end{center}
\caption
{The power spectrum of the primordial potential obtained from a single realization of $\Phi_\vk$ for different choices of $B$ used in the ansatz Eq.~(\ref{eq:discrete_ansatz}). The potential field $\Phi_\vk$ was realized on a grid of size $256^3$ in a $(600\,{\rm Mpc}/h)^3$ box. The dotted black lines show the theoretical predictions. For ``local b)'' and ``local c)'', the deviations from the Gaussian case originate from the term $\langle \Phi^{NG}\Phi^{NG}\rangle$, while for the orthogonal case the deviations are mainly caused by the term $\langle \Phi^{G}\Phi^{NG}\rangle$.}
\label{fig:P_Phi}
\end{figure}

Since the term $\langle \Phi^{NG} \Phi^{NG}\rangle$ is of second order and the potential is small, one could think that this term is negligible for reasonable values of $\fnl$. However, our formula for $\Phi^{NG}$ involves integrals which ---depending on the bispectrum--- are divergent. In order to explore this further, we derive the general expression for $\langle \Phi^{NG} \Phi^{NG}\rangle$. After applying Eq.~(\ref{eq:ansatz}) for $\Phi^{NG}$, using the definition of the power spectrum, and integrating over the delta function we obtain
\begin{equation}
\langle \Phi^{NG}_\vk \Phi^{NG}_{\bf q}\rangle=\frac{1}{18}\delta^D(\vk+{\bf q})\int{d^3k^\prime \frac{ B^2(k,k^\prime,|\vk+{\bf q}|)}{P(k^\prime)P(|\vk+{\bf q}|)}}\,.
\end{equation}
As a first example let us consider the case in which the bispectrum of the local type (Eq.~\ref{eq:B_loc}) is used for $B$ in Eq.~(\ref{eq:ansatz}). In this case, we get three terms

\begin{eqnarray}
\langle \Phi^{NG}_\vk \Phi^{NG}_{\bf q}\rangle=&\frac{2}{9}\fnl^2\delta^D(\vk+{\bf q})\left\{\int{d^3k^\prime P(k^\prime)P(|\vk +{\bf k^\prime}|) } \right. \nonumber \\ 
&+ 4 P(k)\int{ d^3k^\prime P(k^\prime)} \nonumber \\
 &\left.+2 P^2(k) \int{ d^3k^\prime \left( \frac {P(|\vk +{\bf k^\prime}|)}{P(k^\prime)}+1\right)} \right\} \,,
\end{eqnarray}
where the first and second term are very similar to the ones discussed in \cite{Boubekeur2006,McDonald2008}; the second term is just a $k$-independent renormalization, which for reasonable values of $\fnl$ is negligible small because of the truncation of the integral due to the finite volume of the simulation box, while the first term gives rise to a $k$-dependent renormalization which results in a change in the slope of the power spectrum. However, for realistic values of $\fnl$ this change in the slope is very small \cite{McDonald2008}. The last term, proportional to $P^2(k)$, causes the largest modification of the power spectrum. Even for still allowed values of $\fnl$ this term gives significant contributions to the power spectrum on large scales (see Fig.~\ref{fig:P_Phi}). In order to circumvent this problem, we consider $B(k_1,k_2,k_3)\longrightarrow 6\fnl P(k_2)P(k_3)$, which to first order produces the same bispectrum, but does not imply this kind of divergent integral. For this ansatz we obtain only the aforementioned scale-dependent term
\begin{eqnarray}
\langle \Phi^{NG}_\vk \Phi^{NG}_{\bf q}\rangle={2}\fnl^2\delta^D(\vk+{\bf q})\int{d^3k^\prime P(k^\prime)P(|\vk +{\bf k^\prime}|)}  \,.
\end{eqnarray}
In Fig.~\ref{fig:P_Phi}, we show the power spectrum for an extreme $\fnl$ to make the change in the slope visible.

Now we turn to a more general case consisting of combinations of $F^{\rm local}$ (Eq.~\ref{eq:Flocal}), $F^{\rm A}$ and $F^{\rm B}$ (Eq.~\ref{eq:FAB}). To do this we calculate the different products of these three functions. The arguments of the $F$-functions are $k$, $k^\prime$, and $\kappa=|\vk +{\bf k^\prime}|$:

\begin{eqnarray}
\frac{F^{\rm A}F^{\rm A}}{P(k^\prime)P(\kappa)}=P^{4/3}(k)\left[P^{1/3}(k^\prime)P^{1/3}(\kappa)\right]
\end{eqnarray}

\begin{eqnarray}
\frac{F^{\rm A}F^{\rm B}}{P(k^\prime)P(\kappa)}=&P(k)\left[P^{1/3}(k^\prime)P^{2/3}(\kappa)+P^{2/3}(k^\prime)P^{1/3}(\kappa)\right]\nonumber\\
&+P^{4/3}(k)\left[P^{2/3}(\kappa)+P^{2/3}(k^\prime)\right]\nonumber\\
&+P^{5/3}(k)\left[P^{1/3}(\kappa)+P^{1/3}(k^\prime)\right]
\end{eqnarray}

\begin{eqnarray}
\frac{F^{\rm A}F^{\rm local}}{P(k^\prime)P(\kappa)}=&P(k)^{2/3}\left[P^{2/3}(k^\prime)P^{2/3}(\kappa)\right]\nonumber\\
&+P^{5/3}(k)\left[P^{-1/3}(k^\prime)P^{2/3}(\kappa)+P^{2/3}(k^\prime)P^{-1/3}(\kappa)\right]
\end{eqnarray}

\begin{eqnarray}
\frac{F^{\rm B}F^{\rm B}}{P(k^\prime)P(\kappa)}=
&P^{2/3}(k)\left[P^{1/3}(k^\prime)P(\kappa)+P(k^\prime)P^{1/3}(\kappa)+2P^{2/3}(k^\prime)P^{2/3}(\kappa)\right]\nonumber\\
&+P(k)\left[2P(k^\prime)+2P(\kappa)+2P^{2/3}(k^\prime)P^{1/3}(\kappa)+2P^{1/3}(k^\prime)P^{2/3}(\kappa)\right]\nonumber\\
&+P^{4/3}(k)\left[P^{-1/3}(k^\prime)P(\kappa)+P(k^\prime)P^{-1/3}(\kappa)\right.\nonumber\\
&\qquad\qquad\left.+2P^{2/3}(k^\prime)+2P^{2/3}(\kappa)+6P^{1/3}(k^\prime)P^{1/3}(\kappa)\right]\nonumber\\
&+P^{5/3}(k)\left[2P^{-1/3}(k^\prime)P^{2/3}(\kappa)+2P^{2/3}(k^\prime)P^{-1/3}(\kappa)\right.\nonumber\\
&\qquad\qquad\left.+2P^{1/3}(\kappa)+2P^{1/3}(k^\prime)\right]\nonumber\\
&+P^2(k)\left[P^{1/3}(k^\prime)P^{-1/3}(\kappa)+P^{-1/3}(k^\prime)P^{1/3}(\kappa)+2\right]\nonumber\\
\end{eqnarray}

\begin{eqnarray}
\frac{F^{\rm B}F^{\rm local}}{P(k^\prime)P(\kappa)}=
&P^{1/3}(k)\left[P^{2/3}(k^\prime)P(\kappa)+P(k^\prime)P^{2/3}(\kappa)\right]\nonumber\\
&+P^{2/3}(k)\left[P^{1/3}(k^\prime)P(\kappa)+P(k^\prime)P^{1/3}(\kappa)\right]\nonumber\\
&+P(k)\left[P^{2/3}(k^\prime)P^{1/3}(\kappa)+P^{1/3}(k^\prime)P^{2/3}(\kappa)\right]\nonumber\\
&+P^{4/3}(k)\left[P^{-1/3}(k^\prime)P(\kappa)+P(k^\prime)P^{-1/3}(\kappa)\right.\nonumber\\
&\qquad\qquad\left.+P^{2/3}(k^\prime)+P^{2/3}(\kappa)\right]\nonumber\\
&+P^{5/3}(k)\left[P^{-2/3}(k^\prime)P(\kappa)+P(k^\prime)P^{-2/3}(\kappa)\right.\nonumber\\
&\qquad\qquad\left.+P^{1/3}(\kappa)+P^{1/3}(k^\prime)\right]\nonumber\\
&+P^2(k)\left[P^{1/3}(k^\prime)P^{-1/3}(\kappa)+P^{-1/3}(k^\prime)P^{1/3}(\kappa)\right.\nonumber\\
&\qquad\left.+P^{2/3}(k^\prime)P^{-2/3}(\kappa)+P^{-2/3}(k^\prime)P^{2/3}(\kappa)\right]
\end{eqnarray}

\begin{eqnarray}
\frac{F^{\rm local}F^{\rm local}}{P(k^\prime)P(\kappa)}=
&\left[P(k^\prime)P(\kappa)\right]\nonumber\\
&+P(k)\left[2P(k^\prime)+2P(\kappa)\right]\nonumber\\
&+P^2(k)\left[P(k^\prime)P^{-1}(\kappa)+P^{-1}(k^\prime)P(\kappa)+2\right]\nonumber\\
\end{eqnarray}

In the limit of  $k\ll k^\prime$, i.e.~$\kappa\approx k^\prime$, all these second-order terms can be written as $P^r(k)P^{2-r}(k^\prime)$ with $0\le r \le 2$. Using $P(k)\approx k^{-3}$ and performing the truncated integration over $k^\prime$, we find that on large scales the contribution of the terms increase significantly with larger $r$.

For certain combinations of the $F$-functions many of the above terms cancel. For example, with $B\propto (F^{\rm local} - F^{\rm B})$ the $P^2(k)$-terms vanish in the above limit. Moreover for the equilateral case (see Eq.~\ref{eq:Feq}) all terms with $r>1$ cancel, whereas for the enfolded type the term with $P^{4/3}(k)$ does not vanish. 

Now let us return to the term $\langle \Phi^G \Phi^{NG}\rangle$ linear in $\fnl$. In the discretized version this becomes:
\begin{equation}
\langle \Phi^{*G}_\vk\Phi^{NG}_\vk \rangle= \frac {1}{6} \sum_{k^\prime}{\frac{B(k,k^\prime,|\vk+{\bf k^\prime}|)}{P(k^\prime)P(|\vk+{\bf k^\prime}|)}\langle \Phi^*_\vk \Phi^*_{\bf k^\prime} \Phi^*_{(\vk+{\bf k^\prime})}\rangle}\,.
\end{equation}
For small $k$, there are only a few modes in the box and the skewness of the realized Gaussian field is not completely vanishing. Depending on the choice for $B$ this noise can be amplified significantly. In particular, for our ansatz of the orthogonal type (Eq.~\ref{eq:Fort}) this contributes substantially to the power spectrum and causes the deviations visible in Fig.~\ref{fig:Pk_IC}.

\section*{References}
\bibliographystyle{JHEP}

\begin{thebibliography}{99}

\bibitem{ABMR} Acquaviva, V., 
Bartolo, N., Matarrese, S., 
\& Riotto, A.\ 2003, Nuclear Physics B, 667, 119 
\bibitem{maldacena:2003} Maldacena, J.\ 2003, Journal 
of High Energy Physics, 5, 13 
\bibitem{BKMR04} {Bartolo}, N., {Komatsu}, E., {Matarrese}, S., \& {Riotto}, A. 2004, Phys. Rep.,
  402, 103

\bibitem{linde/mukhanov:1997}
Linde, A. \& and Mukhanov, V. 1997, PRD, 56, 535
\bibitem{lyth/ungarelli/wands:2003}
Lyth, D.~H., Ungarelli, C. \& Wands, D. 2003, PRD, 67, 023503
\bibitem{babich/creminelli/zaldarriaga:2004} Babich, D., Creminelli, 
P., \& Zaldarriaga, M.\ 2004, Journal of Cosmology and Astro-Particle Physics, 8, 9 
\bibitem{chen/etal:2007} Chen, X., Huang, M.-x., 
Kachru, S., \& Shiu, G.\ 2007, Journal of Cosmology and Astro-Particle Physics, 1, 2 
\bibitem{holman/tolley:2008}
Holman, R. \& Tolley, A.~J. 2008, JCAP, 0805, 001
\bibitem{chen/easther/lim:2007}
Chen, X., Easther, R. \& E.~A. Lim, E.~A. 2007, JCAP, 0706, 023
\bibitem{langlois/etal:2008}
Langlois, D., Renaux-Petel, S., Steer, D.~A. \& Tanaka, T. 2008, PRD, 78,
	063523 
\bibitem{komatsuwhitepaper} Komatsu, E., et al.\ 
2009, astro2010: The Astronomy and Astrophysics Decadal Survey, 2010, 158 
\bibitem{VM09} Verde, L., \& Matarrese S., 2009, ApJL, 706, L91 
\bibitem{senatore/smith/zaldarriaga:2009} Senatore, L., Smith, 
K.~M., \& Zaldarriaga, M.\ 2010, Journal of Cosmology and Astro-Particle Physics, 1, 28 
\bibitem{meerburg} Meerburg, P.~D., van 
der Schaar, J.~P., 
\& Stefano Corasaniti, P.\ 2009, Journal of Cosmology and Astro-Particle Physics, 5, 18 
\bibitem{VWHK00}
{Verde}, L., {Wang}, L., {Heavens}, A.~F., \& {Kamionkowski}, M. 2000, MNRAS,
  313, 141
\bibitem{V01} Verde, L.\ 2001, The Onset of Nonlinearity in Cosmology, 927, 54 
\bibitem{VJKM01} Verde, L., Jimenez, R., 
Kamionkowski, M., \& Matarrese, S.\ 2001, MNRAS, 325, 412 

\bibitem{Loverdeetal07}
{LoVerde}, M., {Miller}, A., {Shandera}, S., \& {Verde}, L. 2007, ArXiv
  e-prints, 711
\bibitem{Sefusatti} Sefusatti, E.,  Liguori, M., Yadav A.P.S., Jackson M. G.,  Pajer E., arXiv:0906.0232
\bibitem{MVJ00} {Matarrese}, S., {Verde}, L., \& {Jimenez}, R. 2000, ApJ, 541, 10

\bibitem{Grossietal09} Grossi, M., Verde, L., 
Carbone, C., Dolag, K., Branchini, E., Iannuzzi, F., Matarrese, S., 
\& Moscardini, L.\ 2009, MNRAS, 398, 321 
\bibitem{Pillepich} Pillepich, A., 
Porciani, C., \& Matarrese, S.\ 2007, ApJ, 662, 1 
\bibitem{KJV09} Kamionkowski, M., 
Verde, L., \& Jimenez, R.\ 2009, Journal of Cosmology and Astro-Particle Physics, 1, 10 
\bibitem{JV09} Jimenez, R., \& Verde, L.\ 2009, PRD, 80, 127302 
\bibitem{Desjacques} Desjacques, V.,
Seljak, U., \& Iliev, I.~T.\ 2009, MNRAS, 396, 85 
\bibitem{verdereview} Verde, L.\ 2010, arXiv:1001.5217, to appear in Advances in Astronomy

\bibitem{DDHS08}
{Dalal}, N., {Dore}, O., {Huterer}, D., \& {Shirokov}, A.  \ 2008, PRD, 77, 123514 
\bibitem{MV08} Matarrese, S., \& Verde, L.\ 2008, ApJL, 677, L77
\bibitem{afshordi/tolley:2008} Afshordi, N., \& Tolley, A.~J.\  2008, PRD, 78, 123507 
eprint arXiv:0806.1046 
\bibitem{slosar/etal:2008} Slosar, A., Hirata, C., 
Seljak, U., Ho, S., \& Padmanabhan, N.\ 2008, 2008, JCAP, 08, 031.
\bibitem{xiaNVSS10} Xia, J.-Q., Viel, M., 
Baccigalupi, C., De Zotti, G., Matarrese, S., 
\& Verde, L.\ 2010, arXiv:1003.3451 
\bibitem{CVM08} Carbone, C., Verde, L., 
\& Matarrese, S.\ 2008, ApJL 684, L1 
\bibitem{CMV10} Carbone, C., Mena, O., 
\& Verde, L.\ 2010, arXiv:1003.0456 

\bibitem{smith/zaldarriaga:2006} Smith, K.~M., \& Zaldarriaga, M.\ 2006, arXiv:astro-ph/0612571 
\bibitem{fergusson/liguori/shellard:2010} Fergusson, J.~R., 
Liguori, M., \& Shellard, E.~P.~S.\ 2009, arXiv:0912.5516 


\bibitem{Robinson} Robinson, J., \& Baker, J.~E.\ 2000, MNRAS, 311, 781 
\bibitem{Scoccimarro} Scoccimarro, R.\ 2000, ApJ, 542, 1 

\bibitem{vio1} Vio, R., Andreani, P., 
\& Wamsteker, W.\ 2001, PASP, 113, 1009 
\bibitem{vio2} Vio, R., Andreani, P., 
Tenorio, L., \& Wamsteker, W.\ 2002, PASP, 114, 1281 


\bibitem{Salopekbond90}
D.~S. {Salopek} and J.~R. {Bond} PRD, 42:3936--3962, December 1990.

\bibitem{KS01}
E.~{Komatsu} and D.~N. {Spergel} PRD, 63(6):063002--+, March 2001.
\bibitem{seery/lidsey:2005} D. Seery and J. E. Lidsey, Primordial non-Gaussianities in single field inflation, JCAP 
0506, 003 (2005) [arXiv:astro-ph/0503692]. 
\bibitem{khoury/piazza:2009} J. Khoury and F. Piazza, 2009, arXiv:0811.3633 [hep-th]
\bibitem{chen:2005} Chen, X., Phys. Rev. D, 71 (2005) 063506 
[arXiv:hep-th/0408084]
\bibitem{silverstein/tong:2004} Silverstein E., \& Tong, D.  Phys. Rev. D, 70 (2004) 103505 [arXiv:hep-th/0310221]. 


\bibitem{CWinprep} Wagner et al. 2010,  in preparation.

\bibitem{CAMB} Lewis, A., Challinor, A., 
\& Lasenby, A.\ 2000, ApJ, 538, 473 

\bibitem{sirko2005} Sirko, E.\ 2005, ApJ, 634, 728 

\bibitem{McDonald2006} McDonald, P., Trac, 
H., \& Contaldi, C.\ 2006, MNRAS, 366, 547 


\bibitem{springel2005} Springel, V.\ 2005, MNRAS, 
364, 1105 

\bibitem{WMAP7} Komatsu, E., et al.\ 
2010, arXiv:1001.4538 


\bibitem{Gill2004} Gill, S.~P.~D., Knebe, A., 
\& Gibson, B.~K.\ 2004, MNRAS, 351, 399 

\bibitem{knollmann2009} Knollmann, S.~R., \& Knebe, A.\ 2009, ApJS, 182, 608 


\bibitem{Huterer/Takada2005} Huterer, D., \& Takada, M.\ 2005, Astroparticle Physics, 23, 369 



\bibitem{Bartolo2010} Bartolo, N., 
Beltr{\'a}n Almeida, J.~P., Matarrese, S., Pietroni, M., 
\& Riotto, A.\ 2010, Journal of Cosmology and Astro-Particle Physics, 3, 11 

\bibitem{fedeli2010} Fedeli, C., \& Moscardini, L.\ 2010, MNRAS, 405, 681 



\bibitem{Lucchin/Matarrese1988} Lucchin, F., \& Matarrese, S.\ 1988, ApJ, 330, 535 

\bibitem{Tinker2008} Tinker, J., Kravtsov, 
A.~V., Klypin, A., Abazajian, K., Warren, M., Yepes, G., Gottl{\"o}ber, S., 
\& Holz, D.~E.\ 2008, ApJ, 688, 709 

\bibitem{Desjacques2010a} Desjacques, V., \& Seljak, U.\ 2010, Phys. Rev. D, 81, 023006 

\bibitem{Desjacques2010b} Desjacques, V., \& Seljak, U.\ 2010, Classical and Quantum Gravity, 27, 124011 

\bibitem{Boubekeur2006} Boubekeur, L., \& Lyth, D.~H.\ 2006, Phys. Rev. D, 73, 021301 

\bibitem{McDonald2008} McDonald, P.\ 2008, Phys. Rev. D, 78, 123519 






\bibitem{FergussonShallard} J. R. Fergusson and E. P. S. Shellard. The shape of primordial non-Gaussianity and the CMB bispectrum. Phys. Rev.,  D80:043510, 2009. 
\bibitem{Chen} Chen, X.  2010, special issue of Advances in Astronomy, arXiv:1002.1416
\bibitem{strings} Hindmarsh et al, 2010, Phys.Rev.D81:063505
\end{thebibliography}
\providecommand{\href}[2]{#2}\begingroup\raggedright

\end{document}